\documentclass[11pt]{article}
\usepackage{fix-cm}
\PassOptionsToPackage{sort}{natbib}
\usepackage{arxiv}
\usepackage[utf8]{inputenc}
\usepackage[T1]{fontenc}
\usepackage{amsmath,amssymb,amsthm,amsfonts,bm,bbm}
\usepackage{graphicx,booktabs,longtable,enumitem,etoolbox}
\usepackage{xcolor}
\usepackage[labelfont=bf]{caption}
\usepackage{microtype}
\usepackage{nicefrac}
\usepackage{subcaption}
\usepackage{multirow}
\usepackage{float}
\usepackage{placeins}
\usepackage{needspace}
\usepackage{adjustbox}
\usepackage{tikz}
\usepackage[normalem]{ulem}

\colorlet{introeditred}{black}

\newcommand{\regressiontableformat}{\fontsize{9}{11}\selectfont\setlength{\tabcolsep}{2pt}\renewcommand{\arraystretch}{1.15}}
\newcommand{\TVm}{d_{\mathrm{TV}}}

\newcommand{\Hm}{H}

\definecolor{citationblue}{HTML}{03B1F1}
\hypersetup{colorlinks=true,allcolors=citationblue}
\setcitestyle{numbers,square,comma}
\DeclareCaptionFont{papercaption}{\fontsize{10}{12}\selectfont}
\DeclareCaptionLabelSeparator{comma}{,\space}

\AtBeginEnvironment{longtable}{\small\setlength{\tabcolsep}{3pt}}
\allowdisplaybreaks

\colorlet{stimnavy}{black}
\newcommand{\pfield}[1]{{\scshape #1}}
\newcommand{\ptype}[1]{\multicolumn{2}{@{}l@{}}{\textbf{#1}}\\[1.5pt]}
\newcommand{\ptypeg}[2]{\multicolumn{2}{@{}p{\linewidth}@{}}{\textbf{#1}\hspace{7pt}{\itshape #2}}\\[1.5pt]}

\makeatletter
\patchcmd{\l@section}{\setlength\@tempdima{1.5em}}{\setlength\@tempdima{2.3em}}{}{}
\patchcmd{\l@subsection}{2.3em}{3.2em}{}{}
\makeatother
 \usepackage{chapterbib}
\title{Investigating Human--AI Discrepancies via Multiple-Solution Problems}
\author{Zihao Wang\thanks{Department of Mathematics, Stanford University}
\and Francesco Insulla\thanks{Institute of Computational and Mathematical Engineering, Stanford University}
\and 
Andrea Montanari\thanks{Department of Statistics and Department of Mathematics, Stanford University}}

\begin{document}
\raggedbottom
\pdfbookmark[0]{Expanded manuscript}{expanded-manuscript}
\clearpage{}
\begin{cbunit}

\maketitle
\begin{abstract}
  Frontier artificial intelligence (AI) models are benchmarked on whether they reach a correct answer. Yet many problems admit several correct answers and repeated attempts—by different people or by the same model resampled—trace out a distribution over them.
  
  In this work, we ask whether human and model reasoning lead to different distributions over valid solutions. Our testbed comprises 270 reasoning puzzles across five puzzle families. These multiple-solution puzzles each have 3 to 8 valid solutions and are simple enough that humans and models can solve them reliably.  The resulting distributions differ markedly: models differ from one another, yet resemble each other far more than they resemble humans. Model distributions are, moreover,  within every puzzle family, less diverse than human ones. We compare these discrepancies across puzzle categories, and trace how they respond to reasoning-effort settings, to prompting, and to perturbations of the puzzle that leave its solutions unchanged.
  
  Together, these results point at significant differences between human and AI problem-solving processes,
  and their choice among equally defensible solutions.
  As progressive deployment of AI systems in society comes into focus, evaluating such differences 
  (beyond one-dimensional accuracy metrics) is increasingly important. 

  Data and code are available at \url{https://hai-discrepancies.github.io/}.
\end{abstract}

\clearpage
\begingroup
\setcounter{tocdepth}{2}
\setlength{\parskip}{1pt}
\hypersetup{linkcolor=black}
\makeatletter
\let\contentscitationcontext\the@ipfilectr
\tableofcontents
\global\let\the@ipfilectr\contentscitationcontext
\makeatother
\endgroup
\clearpage

\section{Introduction}\label{sec:intro}
There is ample evidence of frontier artificial intelligence (AI) systems
reaching or surpassing human abilities on a variety of problem-solving tasks.
Only in the context of mathematics, humans and AI have been compared in mathematical 
competition problems \cite{balunovic2025matharena,matharena2025putnam},
college level problem-solving \cite{wang2024scibench,liu2024mathbench},
specialized benchmark problems written by professional mathematicians \cite{glazer2024frontiermath},
abstract and visual reasoning \cite{chollet2025arcagi2,mayer2025ivispar,kamradt2026astra},
as well as research-level questions \cite{nie2025uqassessinglanguagemodels,riemannbench2026,hle2026nature,abouzaid2026firstproof}.
Recent AI breakthroughs on long-standing open problems in mathematics 
\cite{openai2026unitdistances,alpoge2026boussinesq,openai2026navierstokes}
indicate that the latest AI systems are comparable to highly skilled human researchers.

This study aims at gaining a broader perspective on the problem-solving behavior
of AI as compared to humans. 
Our starting point is the recognition that most real world problems have multiple solutions 
(or multiple paths to a solution). From routing traffic to designing a piece of hardware, 
it is hard to find examples in which there is a unique valid solution.
Rather than focusing on the one-dimensional summarization
provided by the fraction of correct solutions obtained, we propose
to analyze the distribution over solutions generated by either AI or humans.
Human problem-solving research has studied solution-selection 
in the past
\cite{luchins1942mechanization,zhang1994representations,behrens2023sudoku,leikin2007multiple}, 
but the focus has been on specific cognitive biases, and no comparison was made with AI in the past.

We designed and implemented 270 reasoning puzzles: 100 puzzles in the main
module and 170 in perturbation modules.
Each puzzle has 3 to 8 correct answers and is simple enough
that humans and models can solve it reliably. We posed these puzzles to 521
participants recruited on Prolific, and a hundred times each to each of 
three AI models, see Fig.~\ref{fig:procedure}.
\par

\begin{figure}[!htbp]

\centering\includegraphics[width=\linewidth]{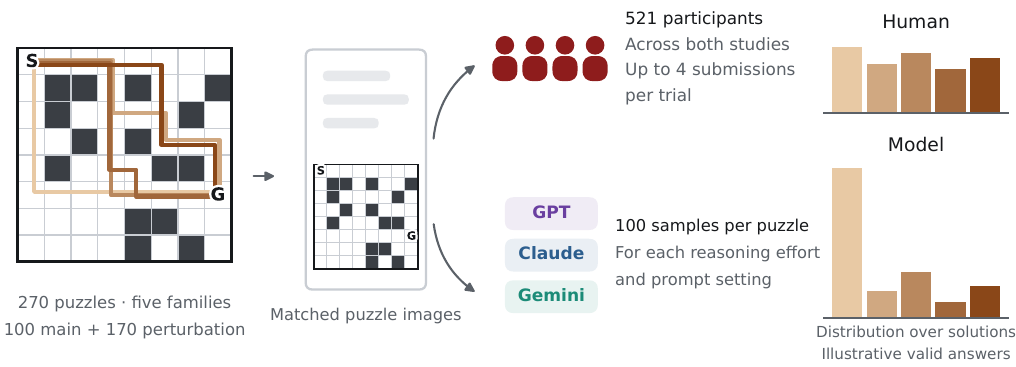}
 \caption{\textbf{General experimental procedure.} Puzzles with multiple valid
solutions are presented to humans and models, and their answer distributions
are compared.}
\label{fig:procedure}
\end{figure}

\FloatBarrier

We find that the distribution of solutions obtained by humans is significantly different from the ones of solutions obtained by three different AI models (GPT 5.6 Sol, Claude Opus 4.8, and Gemini 3.5 Flash), and that different AIs are much closer to each other than they are to humans. In particular, humans return solutions that are much closer to uniformly distributed (hence more diverse) than solutions proposed by AIs, and this distribution is less sensitive to perturbations in the puzzle presentation. On the other hand, the distribution of human solutions is more influenced by previously solved puzzles.
\par

Our findings suggest that different problem solving strategies are followed by humans and AI models, and motivate future work to infer or model these strategies. Also, our study focuses on simple puzzles, and it would be important to understand whether its conclusions extend to problem solving in real-world settings, e.g. medical diagnostics \cite{goh2024diagnostic} or scientific research \cite{luo2025brainbench}. More broadly, understanding differences and complementarities in human-AI reasoning is important in view of the expected societal impact of large scale AI deployment.
\par
\section{Results}\label{sec:results}

We used puzzles of five types, Arithmetic, Maze, Rooks, Minesweeper, and Sudoku.
In the main module, 100 independent puzzles (20 of each type)  are presented to the AI or participant. 
The perturbation modules use five types of perturbations
(each defining a module):
Highlighting, Spatial reflection, Number order, Preceding-puzzle context,
Strategy primer. A  perturbation module comprised $10$ puzzles, plus their variants
(keeping the solution set unchanged).

\subsection{Main module: Distributions over correct answers}\label{sec:results-similarity}

We observe that the distribution of Human
solutions is closer to Uniform than any AI model, while models are
generally closer to one another than to Human. This second ordering is
clearest in Rooks, Minesweeper, and Sudoku.

We measure distance between distributions by total variation (TV). 
Figure~\ref{fig:tv-source-geometry} gives the TV distance
matrix between Uniform, Human, and AI  models, averaged over problem class.  
We compute two-dimensional embeddings of sources by applying 
standard multi-dimensional scaling to the entrywise square 
of the $5\times 5$ TV distance matrix.
This provides a visualization of the  distances between solvers
which explains 80.1\% of the square distances. We observe that the main direction 
separates Human and Uniform from AI models, while the second separates 
GPT from Claude and Gemini, possibly pointing at discrepancies between solution strategies there. 

To visualize distances between puzzle families, we use Centered Kernel Alignment.
We observe that Rooks, Sudoku, and Minesweeper cluster, while  Arithmetic and Maze appear to be more distinct. 
The source and puzzle-family embeddings are described in Methods,
Section~\ref{app:dimension-reduction}.

\begin{figure}[H]
  \centering
  \makebox[\linewidth][c]{\includegraphics[width=183mm]{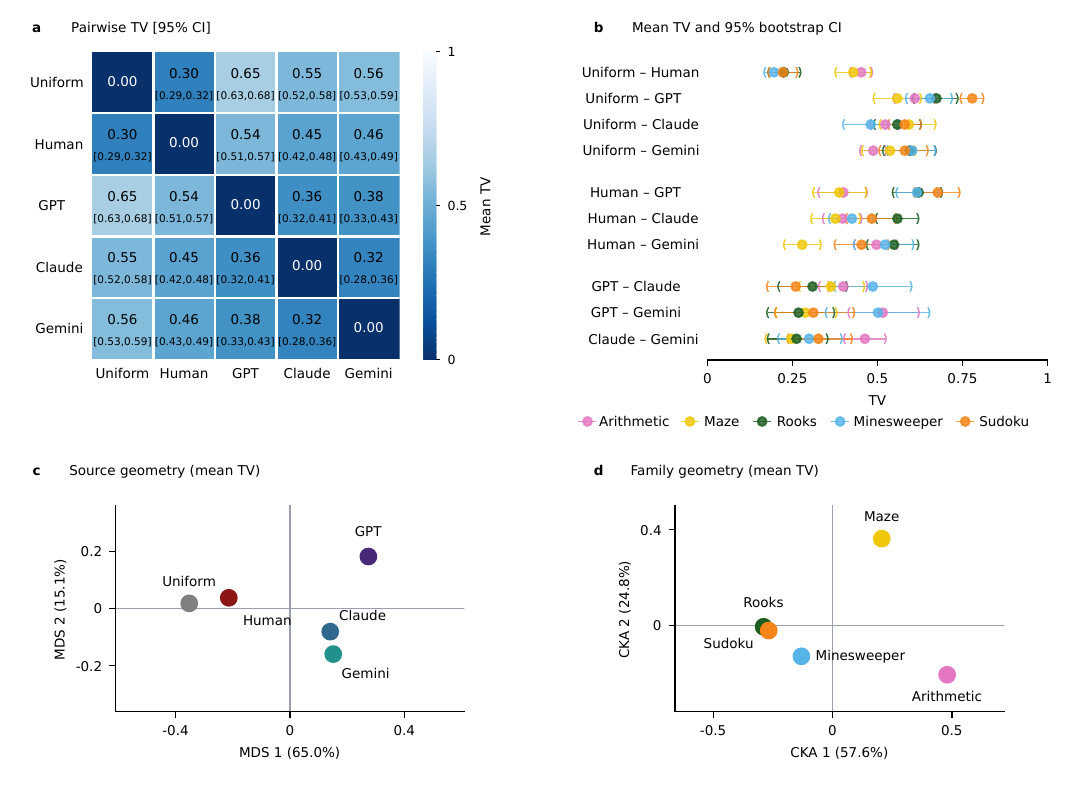}}
  \caption{\textbf{Total-variation distances and source and family geometry.}
  \textbf{a}, Equal-family mean TV distances among Uniform, Human and the three AI
  models, averaged over puzzle family (we report everywhere 
  95\% confidence intervals from 5,000
  whole-puzzle bootstrap resamples within family.)
  \textbf{b},  TV distances by puzzle family, averaged across 20 puzzles per family.
  \textbf{c}, Two-dimensional solver embeddings from multidimensional scaling of
  mean TV across all puzzle families.
  \textbf{d}, Centered kernel alignment
  (CKA) of full positive-semidefinite five-source kernels built from each
  family's mean TV matrix across 20 puzzles.
  Percentages on the axes give the fraction of explained distance. 
  Estimates use 100 main module puzzles (20 per family), conditional
  on correct human or valid model answers; model conditions are low effort
  and plain prompts.}
  \label{fig:tv-source-geometry}
\end{figure}

\Needspace{6\baselineskip}
\subsection{Main module: Entropy}\label{sec:results-entropy}

In order to measure the diversity of solutions produced by each class of solvers, 
we use the normalized Shannon entropy (entropy divided by the maximum entropy, i.e. the logarithm of number of valid solutions per puzzle). 
Normalized entropy equal to zero corresponds to the solver always 
choosing the same solution, while equal to one corresponds to a uniform distribution over
solutions.

Human has the highest mean normalized entropy in every family,
see Table~\ref{tab:normalized-entropy}  and Fig.~\ref{fig:entropy}. Human entropy is lower
in Arithmetic and Maze than in Rooks, Minesweeper, and Sudoku
(and is very close to the maximum value of one in the latter families).
GPT shows the reverse pattern and is nearly deterministic on Sudoku.
Among AI models, we observe a difference between Claude/Gemini and GPT,
as the first two models have higher entropies especially on Sudoku.

\begin{table}[H]
  \color{black}
  \captionsetup{font={color=black}}
  \caption{Normalized Shannon entropy by puzzle family. Same data pipeline as
  for Fig.~\ref{fig:tv-source-geometry}.}
  \label{tab:normalized-entropy}
  \centering
  \setlength{\tabcolsep}{2pt}
  \begin{tabular}{@{}lrrrr@{}}
    \toprule
    & \multicolumn{4}{c}{Mean normalized entropy [95\% CI]} \\
    \cmidrule(l){2-5}
    Puzzle family & Human & GPT & Claude & Gemini \\
    \midrule
    Arithmetic & $0.68[0.65,0.72]$ & $0.45[0.36,0.55]$ & $0.58[0.52,0.64]$ & $0.63[0.59,0.68]$ \\
    Maze & $0.68[0.62,0.74]$ & $0.48[0.38,0.59]$ & $0.39[0.27,0.52]$ & $0.49[0.38,0.61]$ \\
    Rooks & $0.91[0.88,0.94]$ & $0.28[0.18,0.39]$ & $0.47[0.35,0.59]$ & $0.42[0.31,0.53]$ \\
    Minesweeper & $0.93[0.91,0.95]$ & $0.21[0.11,0.32]$ & $0.54[0.43,0.65]$ & $0.33[0.21,0.45]$ \\
    Sudoku & $0.91[0.88,0.94]$ & $0.02[0.01,0.04]$ & $0.42[0.34,0.50]$ & $0.39[0.29,0.51]$ \\
    \midrule
    Mean & $0.82[0.80,0.84]$ & $0.29[0.25,0.33]$ & $0.48[0.43,0.53]$ & $0.45[0.41,0.50]$ \\
    \bottomrule
  \end{tabular}
\end{table}
 
We carried out principal component analysis of the 
entropy table, after subtracting each source's mean across families. The first two singular
directions account for 90.6\% of the remaining variation
(Supplementary Table~\ref{tab:entropy-svd}). 
The resulting picture is consistent with the one obtained from TV distances, cf. 
Fig.~\ref{fig:tv-source-geometry}, (c), (d).
The first component contrasts Arithmetic and Maze with
Rooks, Minesweeper, and Sudoku: Human and GPT lie at opposite ends, Gemini
is on GPT's side, and Claude is near zero.  The second chiefly separates
Maze from Arithmetic and gives Claude the most negative source score.
One possible interpretation is that the first direction reflects the
relative-difficulty proxy discussed in
Section~\ref{sec:results-difficulty}, although with the ordering reversed for
humans. The second may distinguish spatial from symbolic reasoning.

\begin{figure}[!htbp]
\centering
\makebox[\linewidth][c]{\includegraphics[width=175mm]{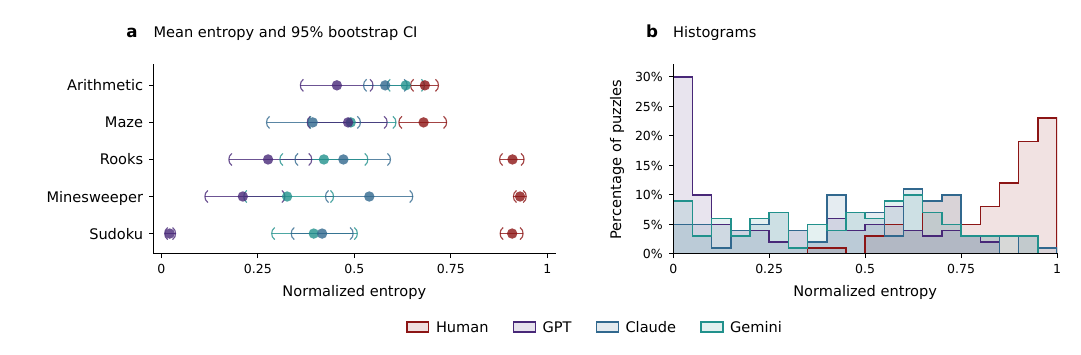}}
\caption{\textbf{Entropy across puzzle families.}
\textbf{a}, Mean normalized Shannon entropy across 20 puzzles per family
and source. 
\textbf{b}, Histograms of normalized entropy for each source over the 100 main
puzzles (20 per family; bins of width 0.05, y axis in percent of puzzles);
ticks below the axis mark the individual puzzles. Same data pipeline as for
Fig. \ref{fig:tv-source-geometry}.}
\label{fig:entropy-profile}
\label{fig:entropy}
\end{figure}

Figure \ref{fig:entropy-profile}, (b) reports the histogram for each solver of normalized entropies (computed over the $100$ puzzles). We observe again that the Human
entropy peaks around the maximum value, while the AI entropies are lower. We also 
observe a marked difference between GPT and the other models.

\subsection{Perturbation modules}\label{sec:results-sensitivity}
Four out of five perturbation modules 
were based on Minesweeper and Sudoku puzzles as follows 
(see Methods, Section~\ref{sec:methods-perturbation-design}).
Highlighting: a subset of cells on the board is highlighted. Spatial reflection: the board is presented in an orientation and three
reflected ones (on different arms of the experiment). Preceding-puzzle context: the solver is asked to solve another
puzzle (either related or unrelated) prior to the current one.
Strategy primer: a primer puzzle is provided alongside the puzzle to be solved.

One perturbation module used Arithmetic puzzles.
Number ordering: the input numbers were presented in either descending or ascending order.

Figure~\ref{fig:presentation} reports observed changes in the probability distribution over solutions induced by these perturbations.
In frame (a), for each of the five perturbations, we report the TV distance
from the reference (unperturbed) distribution (for Preceding-puzzle context,
we use related puzzle pairs and the reference is instead the product of
their observed answer marginals).
In frame (b), we report the normalized change in the probability 
that the solution belongs to the target group (the group that 
is expected to be favored by the perturbation, which for Preceding-puzzle context
consists of answer pairs in which the target-puzzle answer is the mapped
counterpart of the preceding-puzzle answer).
In frame (c), we report the mutual information (MI) between
the solutions of the two puzzles in the Preceding-puzzle context
perturbation. This measures dependence between answers to the preceding and
target puzzles in related and unrelated contexts. Supplementary Section~\ref{supp:perturbation-study}
reports these metrics disaggregated by puzzle family. 

\begin{figure}[!htbp]
\centering\includegraphics[width=\linewidth]{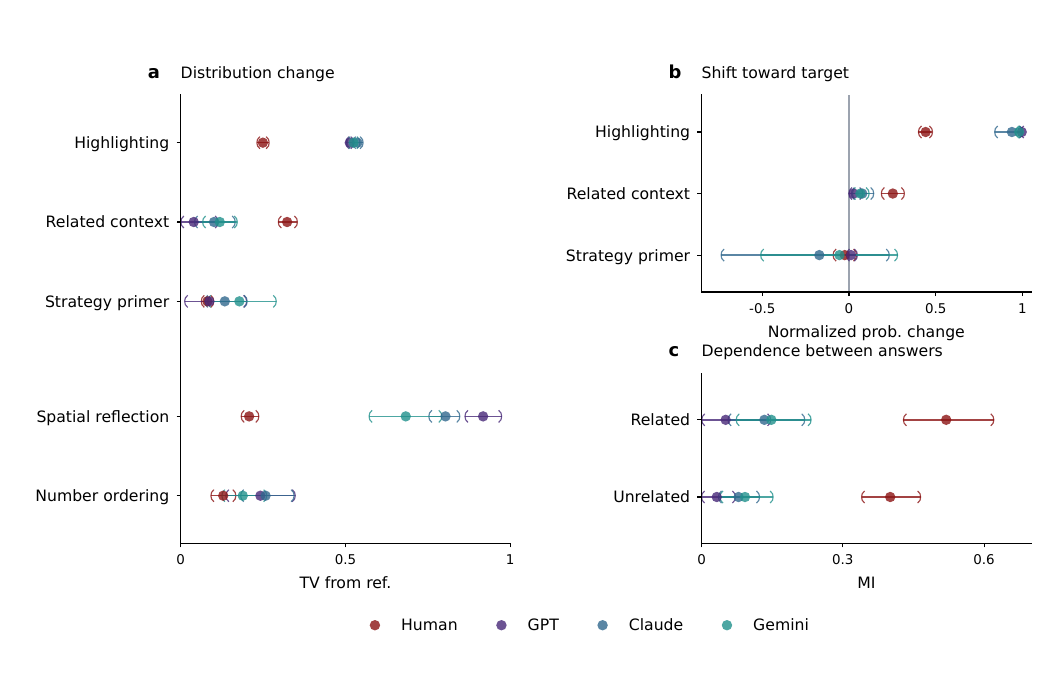}
\caption{\textbf{Effects of presentation and immediate context.}
\textbf{a}, TV from reference. (As a reference distribution: 
Preceding-puzzle context uses the product of its observed
answer marginals for related puzzle pairs; other modules use the unperturbed
answer distribution, after alignment for Spatial reflection.)
\textbf{b}, Normalized probability change for Highlighting, Preceding-puzzle context
and Strategy primer (defined in Section~\ref{sec:target-methods}).
For Preceding-puzzle context, the target group consists of answer pairs in which the
target-puzzle answer is the mapped counterpart of the preceding-puzzle answer.
\textbf{c}, Mutual information (MI), in bits, in related and unrelated context.
Context estimates include only pairs in which both puzzles were answered correctly.}
\label{fig:presentation}
\end{figure}

The effect of perturbations is markedly different in Human and AI
models. AI models are highly sensitive to Highlighting and Spatial reflection, 
despite the fact that these are purely visual changes in the puzzle presentation,
and are essentially insensitive to other perturbations.
Human presents some sensitivity to Highlighting (lower than for AI) and
to Preceding-puzzle context.
In particular, frame (c) shows that Human solutions for two puzzles presented in sequence have significantly larger statistical MI
than AI solutions (the MI in the latter is small, see also Supplementary for tests of independence).

\subsection{Relative difficulty within and between puzzle families}\label{sec:results-difficulty}

We use response time and reasoning-token count as proxies for relative puzzle
difficulty within each source. For humans, we take the median response time
among correct answers; for AI models, the median reasoning-token count among
valid answers. We take natural logarithms and subtract each source's mean
across the 100 main puzzles. These relative-difficulty scores are distinct
from the provider-defined reasoning-effort setting (low or medium).
Figure~\ref{fig:difficulty} displays the scores standardized within each source
to mean zero and sample standard deviation one.

We observe that AI models' relative-difficulty scores are strongly correlated.
In contrast, Human and AI relative-difficulty scores are positively correlated
within families, but negatively correlated overall.
This reflects an underlying complementarity. Arithmetic
has the highest mean relative difficulty for humans but the lowest for every
model. Maze has the highest for models and is below average for humans.
Minesweeper is also below the human average but above average for GPT
and Claude.

\begin{figure}[!htbp]
\centering
\includegraphics[width=\linewidth]{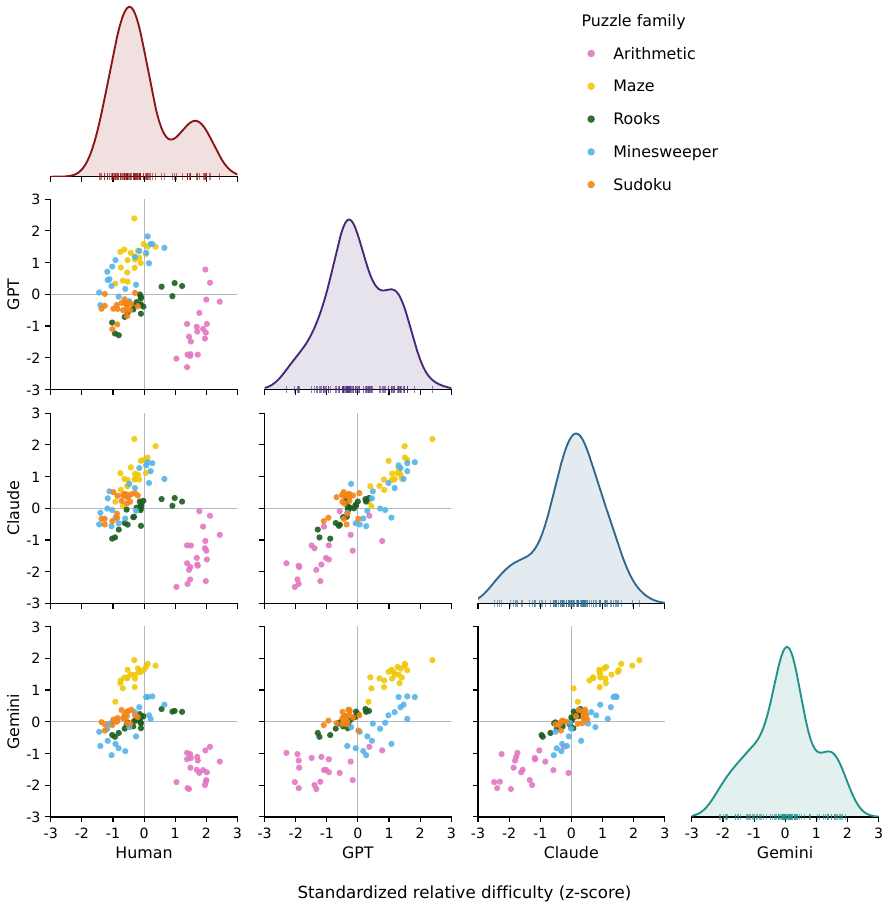}
\caption{\textbf{Relative difficulty across sources.} Relative-difficulty scores for
100 main puzzles (20 per family), standardized within each source to mean
zero and sample standard deviation one. Diagonal panels show kernel density
estimates of each source's scores; ticks mark the individual puzzles.}
\label{fig:effort-correlation}
\label{fig:difficulty}
\end{figure}

\begin{figure}[!htbp]
  \centering
  \includegraphics[width=\linewidth]{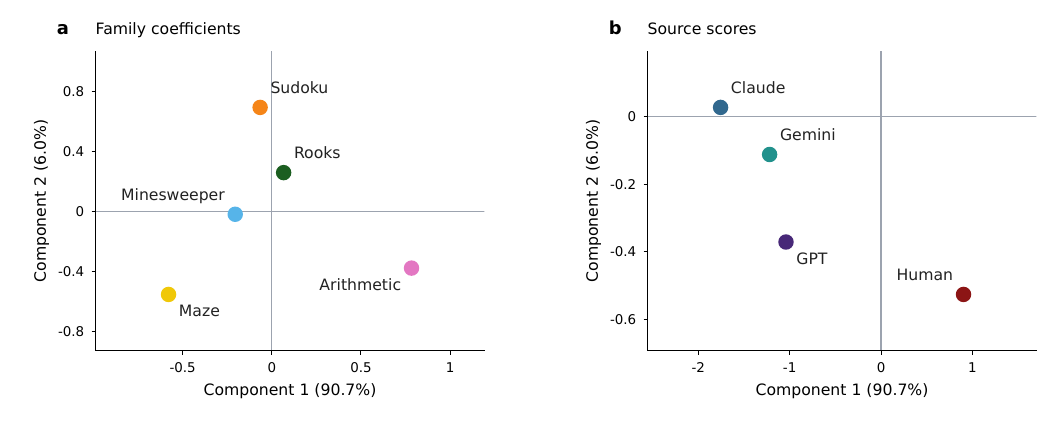}
  \caption{\textbf{Relative-difficulty profile decomposition.} \textbf{a},
  Family coefficients of the first two singular directions. \textbf{b},
  Corresponding source scores. Percentages indicate the share of variation
  in the centered family profiles captured by each direction.}
  \label{fig:effort-svd}
\end{figure}

\Needspace{5\baselineskip}
Figure~\ref{fig:effort-svd} reports the principal component analysis of the
family-by-source matrix of mean relative-difficulty scores.
The first two singular directions capture 96.7\% of variation in these
profiles (Figure~\ref{fig:effort-svd}). The dominant direction separates
Arithmetic from Maze and places humans opposite the models; the second
mainly distinguishes Sudoku.

As mentioned, within a family Human and AI models tend to agree about which
puzzles have higher relative difficulty. Mean within-family Human--model correlations
range from 0.59 to 0.72, similar to the model--model range of 0.60--0.72
(Supplementary Table~\ref{tab:effort-correlation}).
Pooling all 100 puzzles reverses the Human--model association: correlations
range from $-0.51$ to $-0.32$, while model--model correlations remain positive
at 0.76--0.88 (Figure~\ref{fig:effort-correlation}). This reversal reflects
the differences between family profiles, particularly Arithmetic and Maze.

\subsection{Effects of prompting and reasoning-effort settings}\label{sec:results-conditions}

In querying AI models, we vary the prompt (plain or human persona) and reasoning-effort setting (low or
medium), comparing these four conditions within the same puzzles.
Figure~\ref{fig:conditions} shows the TV distances from the Human distribution,
their positions in the source embedding, and their distances from each
model's low plain condition.

\begin{figure}[!htbp]
\centering\includegraphics[width=\linewidth]{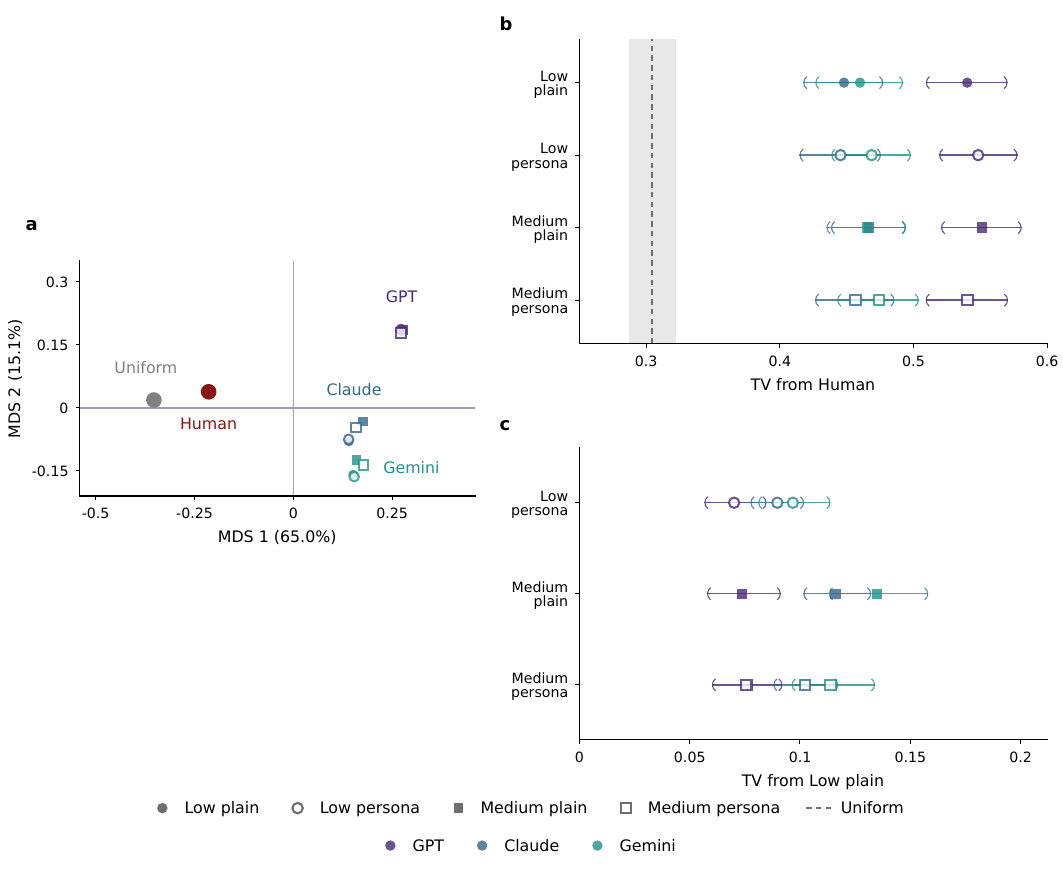}
\caption{\textbf{Distributional distances across reasoning-effort settings and prompts.}
\textbf{a}, Model conditions projected onto the fixed MDS axes
from Figure~\ref{fig:tv-source-geometry}c, using distances to Human, Uniform
and the three Low plain model distributions, see
Section~\ref{app:dimension-reduction}.
\textbf{b}, Mean TV from Human. As a reference, distance from Uniform 
is shown by a dashed
line and band. 
\textbf{c}, Mean TV from each model's Low plain distribution to its other
conditions. In \textbf{b} and \textbf{c}, estimates average 100 main puzzles
(20 per family); bars and the Uniform band are 95\% whole-puzzle bootstrap
intervals from 5,000 resamples within family. Circles indicate the low
reasoning-effort setting and squares the medium setting; filled markers indicate plain prompts and hollow
markers human-persona prompts.}
\label{fig:conditions}
\end{figure}

\FloatBarrier

Neither persona prompting nor a higher reasoning-effort setting consistently brings
AI models' distributions closer to the Human distribution. 
At the medium setting, persona prompting brings
GPT slightly closer to humans, whereas changing Claude's setting from low
to medium under the plain prompt moves it slightly farther away.

Neither intervention consistently increases solution diversity. Mean entropy
generally decreases, but the family effects can oppose one another. Changing
Claude's reasoning-effort setting from low to medium under the plain prompt
broadens the Maze distribution while concentrating the Sudoku distribution;
adding the persona prompt at the medium setting broadens Sudoku again.
\section{Discussion and limitations}\label{sec:discussion}

Correctness or accuracy metrics provide limited insights into the behavior
of AI systems. We propose to complement these assessments by characterizing
the distribution of answers in multiple-solution settings.
This can help predict the behavior of populations of AI agents, 
shed light on the reasoning mechanisms, inform system design or highlight complementarities between AI 
systems and humans.

Our main finding is that frontier models of common use present significantly
less diversity in the choice of solutions than humans. Further, they are
generally more sensitive to changes of the problem presentation that do not change the 
solution space. Finally, relative-difficulty profiles do not necessarily align
between AI models and humans, with different puzzle families having the highest
relative difficulty for humans and AI models.

We emphasize that the evidence we presented is limited to separate U.S.-based Prolific cohorts, three time-specific
AI models (GPT-5.6~Sol, Claude Opus~4.8, and
Gemini~3.5 Flash), two provider-defined reasoning-effort settings, one persona prompt,
default decoding, and small enumerable puzzles.
Each human participant solved a large number of puzzles (either main module, or perturbation modules): we did not attempt to model 
possible dependencies across puzzles solved by the same human.
Also, relative difficulty is estimated from different proxies for humans and
AI models: response time and reasoning-token count, respectively.
Supplementary material presents alternative choices for statistical analysis (e.g.
changing the definition of probability distance), showing that our conclusions are robust with respect to these choices.

It would be important to explore how our findings vary for 
other human cohorts, under other classes of multiple-solution problems,
and varying perturbation stimuli.
An interesting question is whether we can identify along which coordinates Human and AI model 
responses differ. Preliminary analysis (cf. Supplementary material) shows some success for
Maze, but the question is broadly open.
\newpage
\section{Methods}

\subsection{Experiment design}

{\color{black}\hypersetup{citecolor=black,urlcolor=black,linkcolor=black}
\begin{figure}[H]
  \centering
  \includegraphics[width=\linewidth]{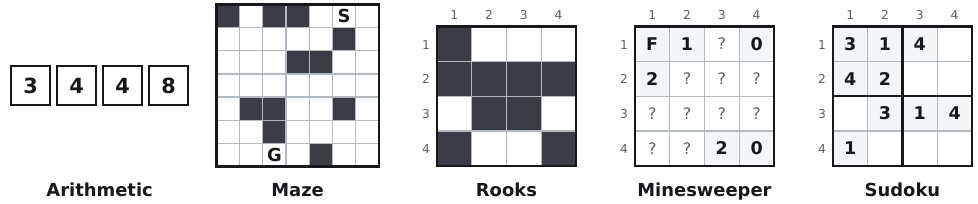}
  
   \caption{\textbf{Examples from the five puzzle families.} One example each
  of Arithmetic, Maze, Rooks, Minesweeper and Sudoku.}
  \label{fig:app-types}
\end{figure}
\par}

\subsubsection{Puzzles}

The five puzzle families are illustrated in Figure
\ref{fig:app-types}: Arithmetic (reach 24 with arithmetic operations on four integers), Maze (shortest path between \texttt{S} and \texttt{G}, 
with up, down, right, left steps allowed), Rooks 
($n-1$ non-attacking rooks on an $n\times n$ board with forbidden cells), Minesweeper (identify a cell that is forced to be a mine or forced to be safe),
and Sudoku (one legal placement). Every puzzle
has between 3 and 8 correct solutions.

\subsubsection{Perturbations}\label{sec:methods-perturbation-design}

Five modules  perturb the stimulus or its
context while preserving the solution set. 
\emph{Highlighting}: we present
each puzzle without highlighting, or highlight one of two complementary,
approximately equal subsets of valid answers together with an equal number
of invalid cells, without changing the clues. For five Sudoku and five
Minesweeper puzzles, these three variants give 30 distinct tasks.
\emph{Spatial reflection}: we present each puzzle in its original form,
reflected left--right, reflected top--bottom, or with both reflections
applied. For five Sudoku and five Minesweeper puzzles, these four variants
give 40 tasks. \emph{Number order}: we
present the same four Arithmetic inputs in nondecreasing order or its exact
reversal, keeping repeated values, operations, the target value and solution
classes unchanged; for ten puzzles, this gives 20 tasks.
\emph{Preceding-puzzle context}: we present each target puzzle alone, paired with a
structurally related predecessor, or paired with an unrelated predecessor
from the same family. For five Sudoku and five Minesweeper target puzzles, this
gives 30 distinct tasks. Related predecessors are transformed versions of
their target puzzles with one-to-one answer maps; unrelated predecessors
are not matched on difficulty, number of valid answers or requested
digit/status. \emph{Strategy primer}: we present each target puzzle alone or paired
with a primer puzzle. For five Sudoku and five Minesweeper
target puzzles, this gives 20 distinct tasks, requesting 30 answers.
The primer puzzle has a unique solution, with a simple local justification;
it must be attempted; no worked solution is supplied. The 
target puzzle is such that one of its answers is prespecified as matching the primer puzzle's local
justification and position pattern. Probability-change estimates
retain valid answers to the target puzzle regardless of success on the primer puzzle. Altogether,
the five modules contain $140$ tasks and $170$ puzzles.

Across modules, each human participant receives one variant per group (e.g. one
out of four variants for Spatial reflection), with assignments
balanced against retained-response quotas and tasks interleaved across
modules. Humans attempt paired puzzles on consecutive paid trials, possibly
separated by attention checks; models see both puzzles side by side in one
image and return both answers in one response. No perturbation puzzle belongs to the main module. However 
seven spatial-reflection puzzles also occur in highlighting or preceding-puzzle context, 
hence a source may solve duplicate puzzles in a session.
Further details are given in Supplementary Section~\ref{app:setup-modules-details}.

\subsubsection{Human participants}

\label{app:setup-human}\label{app:human}
US adults were recruited through Prolific in June 2026. For the main study,
162 participants were recruited, 112 completed the study and 104 were retained.
Each retained participant attempted all 100 main puzzles, yielding 10,400
trials and 9,409 correct responses (90.5\%). The separate perturbation study
comprised 497 recruited and 417 retained participants in two waves, with
25,647 trials and 23,669 correct responses (92.3\%). One person enrolled in
both studies and was excluded from both, giving 658 distinct recruited
participants, 137 excluded and 521 retained. All 417 retained module
participants contributed to all five modules, completing one variant in
each of the 50 design blocks.

Trials had a 120-second limit and permitted up to four answer submissions.
The first three incorrect submissions gave feedback and allowed another
attempt; a fourth incorrect submission automatically ended the trial.
A correct submission ended the trial immediately, and participants could
skip after 90 seconds. Main-study compensation was US\$12 plus US\$0.10 per correct response;
module compensation was US\$9 plus the same bonus. Module distributions use
the final correct answer to a trial. Depending on the module, 11--24\% of
these answers followed at least one rejected submission.

The recorded retention policy excludes testers and requires completion of
all assigned paid responses. Main-study retention permits at most 30 records flagged
as bad overall and at most 14 in any family. The rule uses the recorded
bad-response flag, falling back to timeout, give-up or skip indicators when
that flag is absent. Module retention requires at
least 40 correct trials overall and at least two in each of the five modules.

The study protocol was approved by the Stanford IRB  (approval no. 87192), and all participants provided informed consent before participation.

\subsubsection{Models}

We refer to the three evaluated systems as GPT, Claude, and Gemini.  The
corresponding API endpoints were GPT-5.6~Sol, Claude Opus~4.8, and
Gemini~3.5 Flash, respectively.  We evaluate each
at low and medium provider-specific reasoning-effort settings.
For each model source, the condition set $\mathcal{C}_s$ crosses the two effort
levels with two prompts.  The plain prompt asks the model to solve the
puzzle; the persona prompt adds that it should act as a human participant and
answer as a human would.  Apart from those opening instructions, the prompts
share the same constraints and both end by asking for the first correct
answer that comes to mind.  

Models receive the same underlying boards and instructions as humans, with the
sequence-format difference described above, and answer in a single turn with
no tools, retries, or feedback.  We use provider-default sampling,
do not set temperature or top-$p$, and impose an 8{,}192-token output cap.
For each puzzle and condition we collect 100 separately generated responses.  This gives
96{,}000 image-level requests per provider; because pair and transfer images
elicit two answers, scoring produces 108{,}000 response rows per provider.
Across the completed conditions, 98.0--99.9\% of model responses are valid.

\subsubsection{Analysis population}

\label{app:setup-conditioning}
The primary solution distributions use correct Human responses and valid
model responses only.  Incorrect and invalid responses are removed before
computing these distributions or their discrepancy metrics.  The resulting
comparisons measure which correct solution a source selects, not how often
it is correct.  Every quantity is computed separately
by puzzle, source, and condition before aggregation.

Throughout the paper, $c_0=(\text{low effort},\text{plain prompt})$ is the
primary condition for every model.

\subsection{Statistical methods}\label{sec:notation}
\label{app:stimulus-measures}\label{sec:target-methods}
\label{sec:methods-statistics}\label{supp:puzzle-bootstrap}

Let $\mathcal{H}$ denote the set of puzzle families, and index a family by
$h\in\mathcal{H}$.  For each family $h$, let $\mathcal{T}_h$ be its set of
puzzles, with $t\in\mathcal{T}_h$ indexing a particular puzzle. Let $\mathcal{S}$ denote the set of response sources, indexed by
$s\in\mathcal{S}$.  A source identifies either the human population or a
model provider.  Each source $s$ is evaluated under conditions
$c\in\mathcal{C}_s$; for a model, a condition specifies the reasoning-effort
level and prompt, while the human source has a singleton condition set unless
cohorts are distinguished.  We write $[n]=\{1,\ldots,n\}$, and index the
trials for puzzle $t$ and source $s$ by $i\in[n_{ts}]$. Puzzle $t$ has a finite set of abstract correct solution classes $\mathcal{J}_t$, with cardinality $k_t$.  Responses that differ only by a designated equivalence, such as a commutative reordering of the same arithmetic expression, are assigned the same element of $\mathcal{J}_t$.

Let $\mathcal{I}_{tsc}\subseteq[n_{ts}]$ index the retained correct human
responses or valid model responses, and set
$n_{tsc}=|\mathcal{I}_{tsc}|$.  For each $i\in\mathcal{I}_{tsc}$,
$y_{tsci}\in\mathcal{J}_t$ denotes the response to puzzle $t$ from source $s$
under condition $c$ on trial $i$.  The induced empirical solution distribution
is
\begin{equation}
  p_{tsc}(j\mid\mathrm{correct})
  =\frac{1}{n_{tsc}}\sum_{i\in\mathcal{I}_{tsc}}
    \mathbf{1}\{y_{tsci}=j\},
  \qquad j\in\mathcal{J}_t.
  \label{eq:empirical-distribution}
\end{equation}
We abbreviate this conditional distribution as $\boldsymbol{p}_{tsc}$ when the
conditioning event is clear.  Throughout, $p$ denotes an empirical response
probability or distribution, whereas $\widehat p$ is reserved for a
probability produced by a fitted regression model.

\paragraph{Main distributional summaries.}\label{app:core-measures}
Let $p$ and $q$ be two conditional empirical solution distributions on
$\mathcal{J}_t$.  We write $u_t$ for the uniform distribution on
$\mathcal{J}_t$, so that $u_t(j)=1/k_t$.  Total-variation distance is
\begin{equation}
  \TVm(p,q)=\frac{1}{2}\sum_{j\in\mathcal{J}_t}|p(j)-q(j)|,
  \label{eq:tv}
\end{equation}
and the Shannon entropy of one solution distribution is
\begin{equation}
  \Hm(p)=-\sum_{j\in\mathcal{J}_t}p(j)\log_2 p(j),
  \label{eq:entropy}
\end{equation}
with the convention $0\log_2 0=0$.

For a puzzle with $k_t>1$ valid solution classes, a delta distribution has
$H_{\min,t}=0$ and the uniform distribution has
$H_{\max,t}=\log_2 k_t$.  We define normalized entropy by rescaling
to this feasible range:
\begin{equation}
  \widetilde H_t(p)
  =\frac{\Hm(p)-H_{\min,t}}{H_{\max,t}-H_{\min,t}}
  =\frac{\Hm(p)}{\log_2 k_t}.
  \label{eq:normalized-entropy}
\end{equation}

Given two puzzles $t_1$, $t_2$ (the preceding puzzle and the target puzzle), we also compute the joint empirical distribution by the obvious generalization of the above:
\begin{equation}
  p_{t_1t_2sc}(j_1,j_2\mid\mathrm{correct})
  =\frac{1}{n_{t_1t_2sc}}\sum_{i\in\mathcal{I}_{t_1t_2sc}}
    \mathbf{1}\{y_{t_1sci}=j_1, y_{t_2sci}=j_2\},
  \qquad j_1\in\mathcal{J}_{t_1}, j_2\in\mathcal{J}_{t_2}.
\end{equation}
The mutual information is then defined, with $\mathrm{correct}$ denoting correct answers to both puzzles:
\begin{align}
    {\rm MI}(p_{t_1t_2sc}) = \sum_{(j_1,j_2)\in \mathcal{J}_{t_1}\times
    \mathcal{J}_{t_2}} p_{t_1t_2sc}(j_1,j_2\mid\mathrm{correct}) 
    \log\frac{p_{t_1t_2sc}(j_1,j_2\mid\mathrm{correct}) }{p_{t_1sc}(j_1\mid\mathrm{correct}) 
    p_{t_2sc}(j_2\mid\mathrm{correct})}\, .
\end{align}

For reference and variant distributions $p$ and $q$, and target group $S$
described in Section~\ref{sec:results-sensitivity}, the normalized probability change is
\begin{equation}
  \frac{q(S)-p(S)}{1-p(S)},
  \qquad p(S)=\sum_{j\in S}p(j).
  \label{eq:normalized-attraction}
\end{equation}

Normalized changes are calculated separately for each comparison, then
averaged over eligible comparisons within puzzles, puzzles within families,
and families equally. Tasks with zero denominator are excluded only from
normalized probability-change summaries for that source; they remain included
in raw probability-change and TV summaries. Puzzles with no eligible
comparisons are omitted from the normalized average.

\paragraph{Confidence intervals.}
Bootstrap intervals are pointwise 95\% percentile intervals. For main-study
TV, normalized entropy, relative-difficulty summaries and paired condition
contrasts, we use 5,000 resamples of whole puzzles with replacement within
family. Within each puzzle, we keep the set of solutions fixed.
(Resampling those as well leads to negligible changes.) 
Each resample retains
20 puzzles per family, and family means receive equal weight. Paired condition
values stay together within each sampled puzzle. Supplementary
Table~\ref{tab:tv-global} compares intervals from resampling puzzles, trials,
or both for mean TV between sources.

\paragraph{TV kernel and source projections.}
\label{app:tv-geometry}\label{app:dimension-reduction}
For $S$ sources, let $D^h_{ss'}$ be the mean TV distance between sources
$s,s'$ over puzzles in family $h$, and let $D_{ss'}$ be its mean over
families. The distance matrix $\boldsymbol D=[D_{ss'}]$ has zero diagonal.
We define the entrywise square as 
\begin{equation}
  Q_{ss'}=D_{ss'}^2.
  \label{eq:tv-geometry-dissimilarity}
\end{equation}

With $\boldsymbol C_S=\boldsymbol I_S-
S^{-1}\boldsymbol 1_S\boldsymbol 1_S^\top$, the centered kernel is
\begin{equation}
  \boldsymbol K=-\frac12\boldsymbol C_S\boldsymbol Q\boldsymbol C_S.
  \label{eq:tv-source-kernel}
\end{equation}
This is the classical multidimensional-scaling construction.
 We set the  negative
eigenvalues of $\boldsymbol K$  to zero, obtaining $\boldsymbol K_+$, and record the
discarded spectral mass. If its two largest positive eigenvalues are
$\lambda_1,\lambda_2$, with unit eigenvectors
$\boldsymbol v_1,\boldsymbol v_2$, the source coordinates 
(see e.g. Fig \ref{fig:tv-source-geometry}) are
\begin{equation}\label{eq:mds-coordinates}
  \boldsymbol X
  =\bigl[\sqrt{\lambda_1}\boldsymbol v_1,\,
         \sqrt{\lambda_2}\boldsymbol v_2\bigr]\, ,
\end{equation}
with each row corresponding to one source. 

\textbf{Puzzle family geometry.}
For puzzle family $h$, apply the same conversion and centering to
the TV distance matrix 
$\boldsymbol D^h=[D^h_{ss'}]$, obtaining the 
positive-semidefinite kernel $\boldsymbol K_+^h$. Centered kernel alignment (CKA)
compares two nonzero family kernels:
\begin{equation}
  A_{hh'}
  =\frac{\langle\boldsymbol K_+^h,\boldsymbol K_+^{h'}\rangle_F}
    {\lVert\boldsymbol K_+^h\rVert_F
     \lVert\boldsymbol K_+^{h'}\rVert_F}.
  \label{eq:tv-family-cka}
\end{equation}
CKA equals one for proportional nonzero kernels and removes their overall
scale. 

Finally, the family dissimilarities are $Q^F_{hh'}=2(1-A_{hh'})$.
Centering gives the family kernel
\begin{equation}\label{eq:family-kernel}
-\tfrac12\boldsymbol C_{|\mathcal H|}\boldsymbol Q^F\boldsymbol C_{|\mathcal H|}
=\boldsymbol C_{|\mathcal H|}\boldsymbol A\boldsymbol C_{|\mathcal H|},
\end{equation}
to which we apply the same two-dimensional projection. The resulting
plot places families nearby when they induce similar patterns of
relationships among sources.
 \clearpage
\section*{Data availability}
AI models responses are publicly available at \url{https://hai-discrepancies.github.io/}.

\section*{Code availability}

Code is publicly available at \url{https://hai-discrepancies.github.io/}.

\section*{Acknowledgements}
The authors thank Fan Nie for insightful discussions.

\section*{Funding}
This work was supported by research funds from Stanford University.
A.M. was partially supported by the NSF through Award MFAI-2501597.

 
\begingroup\small\raggedright
\newcommand{\etalchar}[1]{$^{#1}$}
\providecommand{\bysame}{\leavevmode\hbox to3em{\hrulefill}\thinspace}
\providecommand{\MR}{\relax\ifhmode\unskip\space\fi MR }
\providecommand{\MRhref}[2]{%
  \href{http://www.ams.org/mathscinet-getitem?mr=#1}{#2}
}
\providecommand{\href}[2]{#2}

\endgroup
\end{cbunit}
 \clearpage{}
\clearpage{}
\begin{cbunit}
\setcounter{section}{0}
\setcounter{figure}{0}
\setcounter{table}{0}
\setcounter{equation}{0}
\renewcommand{\thesection}{S\arabic{section}}
\renewcommand{\thefigure}{S\arabic{figure}}
\renewcommand{\thetable}{S\arabic{table}}
\renewcommand{\theequation}{S\arabic{equation}}
\renewcommand{\theHsection}{\thesection}
\renewcommand{\theHfigure}{\thefigure}
\renewcommand{\theHtable}{\thetable}
\renewcommand{\theHequation}{\theequation}
\captionsetup[figure]{name=Supplementary Figure}
\captionsetup[table]{name=Supplementary Table}
\pdfbookmark[0]{Supplementary information}{supplementary-information}
\label{part:supplement}

\begin{center}
{\Large\bfseries Supplementary information}\par\medskip
Investigating Human--AI Discrepancies via Multiple-Solution Problems
\end{center}
\bigskip
\noindent\textbf{Contents}\par\medskip
\begingroup\small
\setlength{\tabcolsep}{4pt}
\renewcommand{\arraystretch}{1.02}
\begin{tabular*}{\linewidth}{@{}ll@{\extracolsep{\fill}}r@{}}
\textbf{\ref{supp:further-literature}} & \textbf{\hyperref[supp:further-literature]{Further literature}} & \pageref{supp:further-literature} \\
\addlinespace[3pt]
\textbf{\ref{app:experimental-setup}} & \textbf{\hyperref[app:experimental-setup]{Puzzles and data collection}} & \pageref{app:experimental-setup} \\
\ref{app:setup-puzzles} & \hspace{8pt}\hyperref[app:setup-puzzles]{Main puzzles} & \pageref{app:setup-puzzles} \\
\ref{supp:puzzle-perturbations} & \hspace{8pt}\hyperref[supp:puzzle-perturbations]{Puzzle perturbations} & \pageref{supp:puzzle-perturbations} \\
\ref{app:puzzle-design} & \hspace{8pt}\hyperref[app:puzzle-design]{Study materials} & \pageref{app:puzzle-design} \\
\ref{app:exact-prompts} & \hspace{8pt}\hyperref[app:exact-prompts]{Exact model prompts} & \pageref{app:exact-prompts} \\
\ref{supp:sampling-details} & \hspace{8pt}\hyperref[supp:sampling-details]{Sampling configuration and data totals} & \pageref{supp:sampling-details} \\
\addlinespace[3pt]
\textbf{\ref{app:condition-results}} & \textbf{\hyperref[app:condition-results]{Main study}} & \pageref{app:condition-results} \\
\ref{app:tv-analysis} & \hspace{8pt}\hyperref[app:tv-analysis]{Distributional similarity} & \pageref{app:tv-analysis} \\
\ref{app:entropy-analysis} & \hspace{8pt}\hyperref[app:entropy-analysis]{Entropy} & \pageref{app:entropy-analysis} \\
\ref{supp:effort-summaries} & \hspace{8pt}\hyperref[supp:effort-summaries]{Relative difficulty} & \pageref{supp:effort-summaries} \\
\ref{supp:condition-comparisons} & \hspace{8pt}\hyperref[supp:condition-comparisons]{Effects of reasoning-effort settings and prompting} & \pageref{supp:condition-comparisons} \\
\addlinespace[3pt]
\textbf{\ref{sec:results-features}} & \textbf{\hyperref[sec:results-features]{Predictive value of solution features}} & \pageref{sec:results-features} \\
\ref{sec:methods-prediction} & \hspace{8pt}\hyperref[sec:methods-prediction]{Feature-based prediction} & \pageref{sec:methods-prediction} \\
\ref{sec:maze-geometric-features} & \hspace{8pt}\hyperref[sec:maze-geometric-features]{Maze} & \pageref{sec:maze-geometric-features} \\
\ref{supp:feature-results-arithmetic} & \hspace{8pt}\hyperref[supp:feature-results-arithmetic]{Arithmetic} & \pageref{supp:feature-results-arithmetic} \\
\ref{supp:feature-results-rooks} & \hspace{8pt}\hyperref[supp:feature-results-rooks]{Rooks} & \pageref{supp:feature-results-rooks} \\
\ref{supp:feature-results-minesweeper} & \hspace{8pt}\hyperref[supp:feature-results-minesweeper]{Minesweeper} & \pageref{supp:feature-results-minesweeper} \\
\ref{supp:feature-results-sudoku} & \hspace{8pt}\hyperref[supp:feature-results-sudoku]{Sudoku} & \pageref{supp:feature-results-sudoku} \\
\addlinespace[3pt]
\textbf{\ref{supp:perturbation-study}} & \textbf{\hyperref[supp:perturbation-study]{Perturbation study}} & \pageref{supp:perturbation-study} \\
\ref{supp:perturbation-results-highlighting} & \hspace{8pt}\hyperref[supp:perturbation-results-highlighting]{Highlighting} & \pageref{supp:perturbation-results-highlighting} \\
\ref{supp:perturbation-results-reflection} & \hspace{8pt}\hyperref[supp:perturbation-results-reflection]{Spatial reflection} & \pageref{supp:perturbation-results-reflection} \\
\ref{supp:perturbation-results-context} & \hspace{8pt}\hyperref[supp:perturbation-results-context]{Preceding-puzzle context} & \pageref{supp:perturbation-results-context} \\
\ref{supp:perturbation-results-primer} & \hspace{8pt}\hyperref[supp:perturbation-results-primer]{Strategy primer} & \pageref{supp:perturbation-results-primer} \\
\addlinespace[3pt]
\end{tabular*}
\endgroup
\clearpage
\section{Further literature}\label{supp:further-literature}

Human problem-solving research has long treated the selected route or
procedure as data rather than incidental variation \cite{luchins1942mechanization}.
Among others, \cite{bailenson2000initial,tong2022route,lancia2023shortcuts}
study mechanisms and biases in route selection,
\cite{lynch2022sudoku,behrens2023sudoku} analyze strategies
used in solving Sudoku,
and \citep{cheyette2025decompose} studies
Minesweeper.
Some of the perturbations we study have been investigated before,
although with different motivations. Among these are
visual highlighting \citep{grant2003eye} and
the use of analogy with prior examples \citep{gick1980analogical}.

Frontier-model reports summarize reasoning performance mainly through accuracy
or pass rates \cite{singh2025openai,anthropic2026opus48,google2026gemini35flash}. 
Recent reasoning benchmarks likewise make correctness their primary outcome
\citep{rein2024gpqa,hle2026nature,chollet2025arcagi2,jimenez2024swebench}.

The closest precedent for our work is a recent line of research on model behavioral fidelity 
of large language models, with applications to the social sciences.  Argyle
et al.\ use language models to simulate survey samples
\citep{argyle2023outofone}.  Santurkar et al.\ examine which demographic
groups' opinions language models reflect \citep{santurkar2023whose}.  Aher et al.\ use language
models to replicate human-subject studies \citep{aher2023simulating}.  Horton
studies language models as simulated economic agents
\citep{horton2023large}.  Interview-grounded agents have been compared with
repeated responses from the same participants and evaluated across
experimental outcomes \citep{park2024generativeagents}.  

Output diversity and behavioral fidelity are related but not equivalent.
Reinforcement learning from human feedback can reduce diversity relative to
supervised fine-tuning \citep{kirk2024understanding}.
Instruction-tuned models can also remain concentrated when asked to sample
randomly from valid choices, although targeted fine-tuning can spread their
probability mass \citep{zhang2024forcing}.  
\section{Puzzles and data collection}
\label{app:experimental-setup}

\subsection{Main puzzles}
\label{app:setup-puzzles}
\label{app:solution-classes}
\label{app:setup-solvers}
\label{app:solution-class-details}

The main battery contains 100 puzzles, twenty from each of five families,
with three to eight valid solution classes under the equivalences below.
Figure~\ref{fig:app-types} illustrates the five families.
Below we provide a plain english description of the puzzle 
structure, as well as the space of valid solutions (this requires
to define when two solutions are regarded as equivalent, a.k.a.
`canonicalization').

\Needspace{17\baselineskip}
\subsubsection{Arithmetic}
Combine the four displayed positive integers into an expression equal to 24,
using each occurrence exactly once and only binary addition, subtraction,
multiplication or division, with arbitrary parentheses. 
Input integer range from $1$ to $13$ and are displayed in ascending order.
Negative or
fractional intermediate values are allowed; division by zero, concatenation,
and additional constants are forbidden.

Canonicalization recursively flattens addition and subtraction into sorted
multisets of positive and negative terms, and multiplication and division
into sorted numerator and denominator factors. 
Addition and multiplication merge corresponding groups; subtraction and division first swap the right operand’s two groups. Thus $a-(b-c)$ and $a-b+c$ are equivalent, as also $a/(b/c)$ and $(a\cdot c)/b$.
Equal-valued input occurrences are interchangeable.

\subsubsection{Maze}
The board is a square grid with open cells, walls, a start \texttt{S}, and a
goal \texttt{G}. Moves are one cell up, down, left or right and cannot cross
walls or leave the board. A valid answer is a shortest path from start to
goal, encoded by the corresponding letters \texttt{U}, \texttt{D}, \texttt{L}
and \texttt{R}. {Each complete move string defines a distinct solution class.}
Main boards are $6\times6$, $7\times7$ or $8\times8$, with
shortest paths of eight to thirteen moves.

\subsubsection{Rooks}
Place $n-1$ indistinguishable tokens on allowed cells of an $n\times n$ board,
with no two tokens in the same row or column. Main boards use
$n\in\{4,5,6\}$, and cells marked \texttt{X} are forbidden. Diagonal
alignment is allowed. An answer lists the occupied row--column coordinates.
{Listing order is ignored. Different placements, including symmetric ones,
remain distinct solution classes.}

\subsubsection{Minesweeper}
Each number gives the count of mines in neighboring cells, including diagonal
neighbors. A flag marks a known mine, and a question mark marks a hidden cell.
The prompt asks for one hidden cell guaranteed to be safe or to contain a
mine, as specified for that puzzle. This status must hold in every mine
assignment consistent with the visible clues and flags. Main boards are
$4\times4$ or $5\times5$, with seven to fourteen hidden cells.
{Solution classes are the selected coordinates, not the complete mine layouts.}

\subsubsection{Sudoku}
Select an empty cell in which the specified digit can be placed without
repeating it in the cell's row, column or outlined box.
{Each legal coordinate is a separate solution class.}
Main boards are $4\times4$ with $2\times2$ boxes or $6\times6$ with $2\times3$
boxes. Printed digits remain fixed. Validity is local: the placement need
not be forced or extend to a complete Sudoku solution.

\Needspace{8\baselineskip}
\subsection{Puzzle perturbations}\label{supp:puzzle-perturbations}\label{app:setup-modules-details}

In what follows, we will use the terms
`puzzle', `variant', `task', `trial' as follows.
A \emph{puzzle} is a single problem requiring one answer (e.g. a single maze). 
A \emph{variant} specifies a puzzle's presentation or context (e.g. flipped left to right, or with related context). 
A \emph{task} is a distinct puzzle--variant combination, containing either a single puzzle or a pair. 
A \emph{trial} is a presentation of a puzzle to a human participant or of a task to an AI model. Humans attempt paired puzzles sequentially (two trials), whereas AI models receive both puzzles in one request and return two answers (one trial).




\Needspace{8\baselineskip}
\subsubsection{Highlighting}\label{app:design-highlighting-details}
Each base puzzle presents the same board without highlighting and in two highlighting
arms. We divide the valid answer cells into two fixed groups, one per arm,
with each cell belonging to exactly one group. The groups are equal in size
or, when the total is odd, differ by one cell. The boards have five to seven
valid answers, so an arm highlights two to four valid cells.

Each arm adds an equal number of invalid decoy cells, drawn from hidden
cells in Minesweeper or empty cells in Sudoku. The decoy sets are disjoint
between arms. In Minesweeper, a decoy is not a valid answer to the requested
safe-or-mine question; it may be unresolved rather than provably of the
opposite status. In Sudoku, it is an empty cell where the requested digit
would violate a local constraint. All selected cells receive the same gold
outline. Participants are not told which are valid targets and which are
decoys.

\begin{figure}[H]
    
  \centering
  \includegraphics[width=0.80\linewidth]{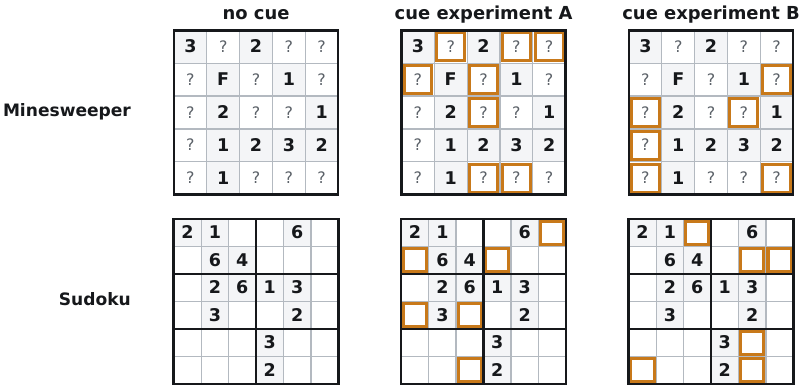}
   \caption{\textbf{Highlighting}: one example puzzle per family. The two highlighting variants
  outline disjoint cells and are the only perturbation to the puzzle.}
  \label{fig:app-cue}
\end{figure}

\Needspace{8\baselineskip}
\subsubsection{Spatial reflection}\label{app:design-reflection-details}
Each base puzzle has four presentations: the original board, its left--right
reflection, its top--bottom reflection, and both reflections together.
The left--right reflection reverses the order of columns; the top--bottom
reflection reverses the order of rows. Applying both is equivalent to
rotating the board by half a turn. Symbols move to their transformed cells
but are redrawn upright; their glyphs are not mirror images.

All five Minesweeper boards are $5\times5$, and all five Sudoku boards are
$6\times6$. The requested safe-or-mine status or Sudoku digit stays fixed.
Responses are mapped back to
the original coordinates before comparing distributions for presentation
invariance.

\begin{figure}[H]
    
  \centering
  \includegraphics[width=0.97\linewidth]{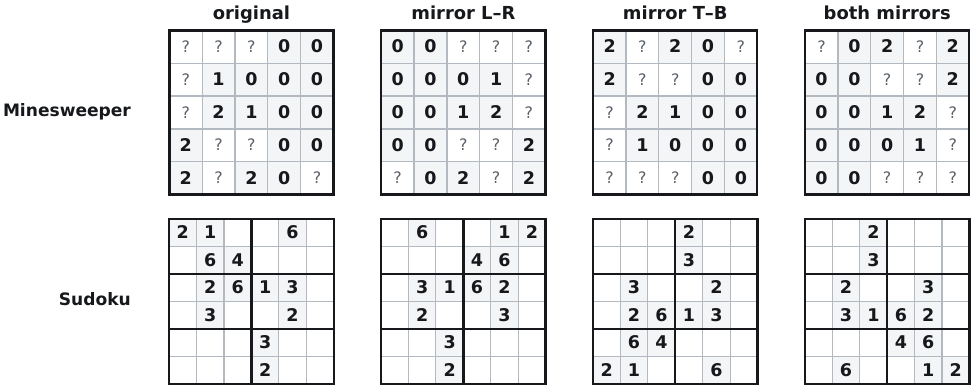}
   \caption{\textbf{Spatial reflection}: a board and three transformed versions,
  preserving solution identities under the corresponding coordinate map.}
  \label{fig:app-spatial}
\end{figure}

\Needspace{8\baselineskip}
\subsubsection{Number ordering}\label{app:design-number-order-details}
Each of the ten Arithmetic puzzles is shown in nondecreasing input order
and its exact reversal, preserving repeated values, operations, target, and
instructions. Both presentations use the same canonical expression catalog
(Section~\ref{app:solution-classes}), so any distributional difference reflects
selection among the same valid classes.

\begin{figure}[H]
    
  \centering
  \includegraphics[width=0.55\linewidth]{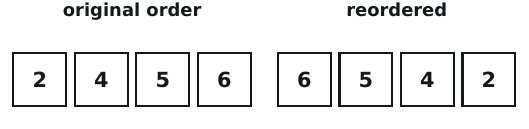}
   \caption{\textbf{Number ordering}: the same four numbers displayed in
  a different order.}
  \label{fig:app-form}
\end{figure}

\Needspace{8\baselineskip}
\subsubsection{Preceding-puzzle context}\label{app:design-context-details}
Each base puzzle has three variants: target puzzle B alone, a related puzzle A
with B, and an unrelated puzzle A with B. The B board, requested answer
type or digit, and valid answer catalog are identical across the three
variants. In the related variant, A and B are two presentations of the
same underlying puzzle with a verified one-to-one answer map. The frozen
Minesweeper pairs use rotations or a reflection. The Sudoku pairs use
reflections or a half-turn, sometimes combined with a row swap within a band
or a column swap within a stack. These transformations preserve the local
placement constraints and the requested digit.

The unrelated A is a different puzzle of the same task family, with no
designated answer correspondence to B. It provides a comparison for the
effect of encountering another puzzle, rather than another presentation of
the same puzzle. It is not matched to related A on difficulty, answer count,
or requested target: all five Sudoku controls request a different digit
from B, and two Minesweeper controls switch between a safe-cell and a
mine-cell question. The control therefore does not isolate relatedness from
every other property of the predecessor.

Joint analyses require two valid answers.

\begin{figure}[H]
    
  \centering
  \includegraphics[width=\linewidth]{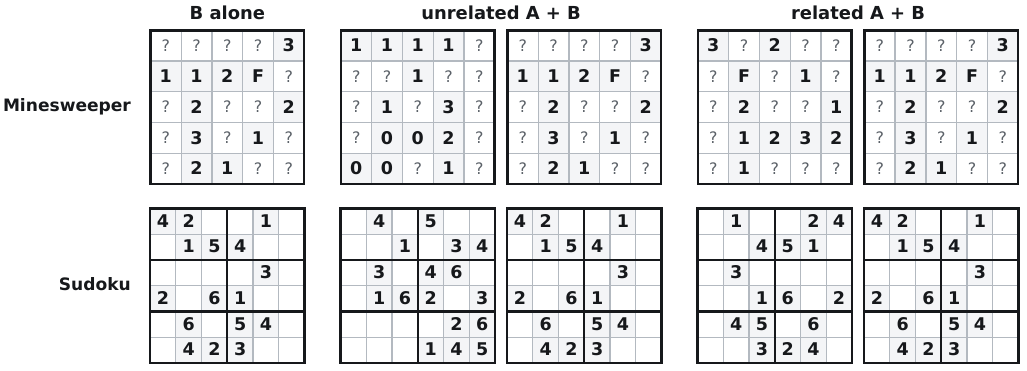}
   \caption{\textbf{Preceding-puzzle context}: the target puzzle B alone, or displayed together
  with an unrelated or a related puzzle A.}
  \label{fig:app-pair}
\end{figure}

\Needspace{8\baselineskip}
\subsubsection{Strategy primer}\label{app:design-primer-details}
We compare each multi-solution target puzzle B alone with the identical B
accompanied by a single-solution primer puzzle A. The primer is an ordinary puzzle,
not a worked example or an explicit instruction to use a strategy. Its
unique answer has a  local justification. One of B's three to
seven valid answers is selected in advance as matching that justification.
Unlike the related-puzzle module, there is no bijection between the answer
sets of A and B.

The justifications for the single solutions are as follows.

Three Minesweeper puzzles use a \emph{satisfied clue}: the adjacent flags
already account for the clue's mine count, so its sole adjacent hidden cell
must be safe. Two use an \emph{exact remaining mine}: after subtracting the
adjacent flags, one mine remains and there is one adjacent hidden cell, so
that cell must be mined. The designated answer in B matches the primer's
local clue, flag, and target-cell pattern up to rotation or reflection.
Other B answers can satisfy the same broad rule through a different local
pattern; the designated target is the pattern match, not every answer to
which the rule could apply.

Three Sudoku puzzles use \emph{row completion}: the designated cell is legal
for the requested digit and is the only empty cell in its row, but not the
only empty cell in its box. Two use \emph{box completion}, reversing these
row and box conditions. These criteria concern empty cells, not the number
of legal locations for the requested digit. The primer puzzle and target puzzle
have the same board size, and the designated cells share either their column
or their position within the box.

The comparison measures probability assigned to B's prespecified matching
cell among valid B answers, regardless of A's correctness
(Methods, Section~\ref{sec:methods-perturbation-design}). 
Because matching combines a local
rule with visual or positional similarity, a shift toward that cell does
not establish transfer of an abstract strategy.

\begin{figure}[H]
    
  \centering
  \includegraphics[width=0.79\linewidth]{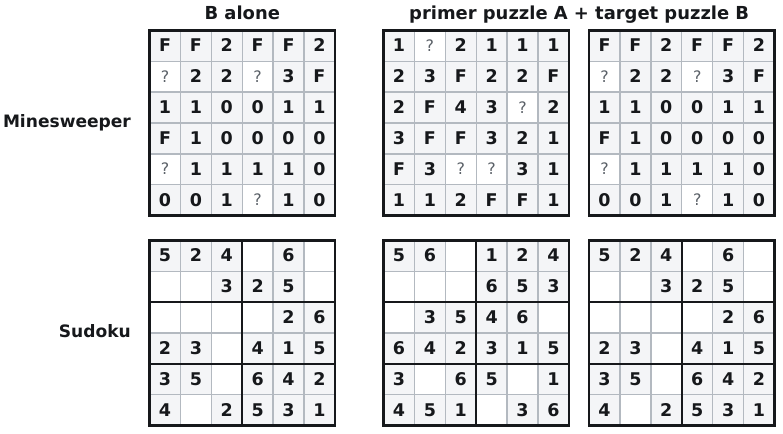}
  
   \caption{\textbf{Strategy primer}: the multi-solution target puzzle B alone, or
  with a single-solution primer puzzle A that can be solved by a designated local
  rule. The primer is an ordinary puzzle, not a worked demonstration. Humans
  see A before B; models receive both panels in one image.}
  \label{fig:app-transfer}
\end{figure}
\clearpage
\subsection{Study materials}\label{app:puzzle-design}
The panels below reproduce the task instructions verbatim, including the
original puzzle-family labels.

\begin{figure}[H]
    
  \centering
  \begingroup\fontfamily{qhv}\selectfont
  \frenchspacing
  \setlength{\tabcolsep}{0pt}
  \renewcommand{\arraystretch}{1.05}
  \begin{tabular}{@{}p{0.56in}p{\dimexpr\linewidth-0.56in\relax}@{}}
    \toprule
    \ptype{Arithmetic}
    \pfield{task}   & Make exactly 24 using the four displayed numbers.\\
    \pfield{rules}  & Use each number exactly once. Allowed operations: +, -,
                      *, /. Parentheses are allowed. Do not use other numbers.
                      Only one correct answer is needed.\\
    \pfield{output} & Return exactly one line: ANSWER: expression=<expression>.
                      Replace placeholders with your answer.\\
    \addlinespace[5pt]
    \ptype{Maze}
    \pfield{task}   & Find one shortest path from S to G.\\
    \pfield{rules}  & Move only U/D/L/R. White cells are open; black cells are
                      walls. S and G are open. The path must be shortest. Only
                      one correct answer is needed.\\
    \pfield{output} & Return exactly one line: ANSWER: path=<UDLR-moves>.
                      Replace placeholders with your answer.\\
    \addlinespace[5pt]
    \ptypeg{Rooks}{$k$ tokens per puzzle, $k\in\{3,4,5\}$; one
      <r,c> slot per token}
    \pfield{task}   & Place $k$ tokens on the board.\\
    \pfield{rules}  & Use row/column numbers. Place exactly $k$ tokens. Do not
                      place on X. No two tokens may share a row or column.
                      Only one correct answer is needed.\\
    \pfield{output} & Return exactly one line: ANSWER: cells=<r1,c1>;
                      <r2,c2>; \ldots; <r$k$,c$k$>. Replace placeholders with
                      your answer.\\
    \addlinespace[5pt]
    \ptypeg{Minesweeper}{brackets: the target cell wording}
    \pfield{task}   & Select exactly one hidden cell marked ?. Each number
                      gives the exact count of mines in the up to eight
                      neighboring cells, including diagonals. F marks a known
                      mine. Your selected ? cell must be forced by the clues:
                      in every mine arrangement consistent with all numbers,
                      that cell is \textbf{safe} [a \textbf{mine}].\\
    \pfield{rules}  & Only one correct answer is needed.\\
    \pfield{output} & Return exactly one line: ANSWER: cell=<row,column>.
                      Replace placeholders with your answer.\\
    \addlinespace[5pt]
    \ptypeg{Sudoku}{X: the target digit, 1--6}
    \pfield{task}   & Select exactly one empty cell for digit \textbf{X}. A
                      cell is legal only if \textbf{X} does not already appear
                      in the same row, the same column, or the same outlined
                      box. Printed numbers are fixed clues and cannot be
                      selected.\\
    \pfield{rules}  & Only one correct answer is needed.\\
    \pfield{output} & Return exactly one line: ANSWER: cell=<row,column>.
                      Replace placeholders with your answer.\\
    \bottomrule
  \end{tabular}

   \endgroup
  \caption{\textbf{Exact single-puzzle instructions.}  Text displayed above
  the boards, reproduced word for word from the rendered stimuli.}
  \label{fig:app-cards}
\end{figure}

\begin{figure}[!htbp]
    
  \centering
  \begingroup\fontfamily{qhv}\selectfont
  \frenchspacing
  \setlength{\tabcolsep}{0pt}
  \renewcommand{\arraystretch}{1.05}
  \begin{tabular}{@{}p{0.56in}p{\dimexpr\linewidth-0.56in\relax}@{}}
    \toprule
    \ptype{Two puzzles: Minesweeper}
    \pfield{task}   & Solve Puzzle A and Puzzle B.\\
    \pfield{rules}  & Use the rule statement shown above each Minesweeper
                      board. Only one correct answer is needed for each
                      puzzle.\\
    \pfield{output} & Return exactly two lines and nothing else. Replace
                      placeholders with your answers.\\
    \pfield{line 1} & A: cell=<row,column>\\
    \pfield{line 2} & B: cell=<row,column>\\
    \addlinespace[3pt]
    \multicolumn{2}{@{}p{\linewidth}@{}}{\textit{Above each board, under
      its Puzzle A / Puzzle B label:}}\\[1pt]
    & Select exactly one hidden cell marked ?. Each number gives the exact
      count of mines in the up to eight neighboring cells, including
      diagonals. F marks a known mine. Your selected ? cell must be forced
      by the clues: in every mine arrangement consistent with all numbers,
      that cell is \textbf{safe} [a \textbf{mine}].\\
    \bottomrule
  \end{tabular}

  \vspace{8pt}
  \begin{tabular}{@{}p{0.56in}p{\dimexpr\linewidth-0.56in\relax}@{}}
    \toprule
    \ptype{Two puzzles: Sudoku}
    \pfield{task}   & Solve Puzzle A and Puzzle B.\\
    \pfield{rules}  & Use the rule statement shown above each Mini Sudoku
                      board. Only one correct answer is needed for each
                      puzzle.\\
    \pfield{output} & Return exactly two lines and nothing else. Replace
                      placeholders with your answers.\\
    \pfield{line 1} & A: cell=<row,column>\\
    \pfield{line 2} & B: cell=<row,column>\\
    \addlinespace[3pt]
    \multicolumn{2}{@{}p{\linewidth}@{}}{\textit{Above each board, under
      its Puzzle A / Puzzle B label:}}\\[1pt]
    & Select exactly one empty cell for digit \textbf{X}. A cell is legal
      only if \textbf{X} does not already appear in the same row, the same
      column, or the same outlined box. Printed numbers are fixed clues and
      cannot be selected.\\
    \bottomrule
  \end{tabular}

   \endgroup
  \caption{\textbf{Exact two-puzzle instructions.}  Text displayed above the
  boards, reproduced word for word from the rendered stimuli.}
  \label{fig:app-cards-two}
\end{figure}

\FloatBarrier
\subsection{Exact model prompts}\label{app:exact-prompts}

\begin{figure}[!htbp]
  \centering
  \begingroup\fontfamily{qhv}\selectfont
  \frenchspacing
  \setlength{\tabcolsep}{0pt}
  \renewcommand{\arraystretch}{1.05}
  \begin{tabular}{@{}p{0.14in}p{\dimexpr\linewidth-0.14in\relax}@{}}
    \toprule
    \ptype{Plain prompt, single-puzzle trials}
    & Solve the puzzle shown in the image.\\
    & Return exactly one final answer line and nothing else.\\
    & Your final line must exactly match the answer format shown in the
      image, including labels, punctuation, separators, and capitalization.\\
    & Replace placeholders with your answer.\\
    & Do not include any explanation, reasoning, preface, markdown, extra
      whitespace, or additional lines.\\
    & Give the first correct answer that comes to mind.\\
    \addlinespace[5pt]
    \ptype{Persona prompt, single-puzzle trials}
    & Act as a human participant solving this task.\\
    & Solve the puzzle shown in the image the way a human participant
      would.\\
    & Return exactly one final answer line and nothing else.\\
    & Your final line must exactly match the answer format shown in the
      image, including labels, punctuation, separators, and capitalization.\\
    & Replace placeholders with your answer.\\
    & Do not include any explanation, reasoning, preface, markdown, extra
      whitespace, or additional lines.\\
    & Give the first correct answer that comes to mind.\\
    \addlinespace[5pt]
    \ptype{Plain prompt, two-puzzle trials}
    & Solve Puzzle A and Puzzle B shown in the image.\\
    & Return exactly two final answer lines and nothing else.\\
    & Each final line must exactly match the corresponding answer format
      shown in the image, including labels, punctuation, separators, and
      capitalization.\\
    & Replace placeholders with your answers.\\
    & Do not include any explanation, reasoning, preface, markdown, extra
      whitespace, or additional lines.\\
    & For each puzzle, give the first correct answer that comes to mind.\\
    \addlinespace[5pt]
    \ptype{Persona prompt, two-puzzle trials}
    & Act as a human participant solving this task.\\
    & Solve Puzzle A and Puzzle B shown in the image the way a human
      participant would.\\
    & Return exactly two final answer lines and nothing else.\\
    & Each final line must exactly match the corresponding answer format
      shown in the image, including labels, punctuation, separators, and
      capitalization.\\
    & Replace placeholders with your answers.\\
    & Do not include any explanation, reasoning, preface, markdown, extra
      whitespace, or additional lines.\\
    & For each puzzle, give the first correct answer that comes to mind.\\
    \bottomrule
  \end{tabular}
  \endgroup
  \caption{\textbf{Exact model prompts.}  The four prompts are reproduced word
  for word.  Every request carries one of them plus the rendered image.}
  \label{fig:app-api-prompt}
\end{figure}

\FloatBarrier

\subsection{Sampling configuration and data totals}\label{supp:sampling-details}\label{app:api}
Tables~\ref{tab:api-config} and~\ref{tab:data} record the sampling
configuration and collection totals.

\FloatBarrier


\begin{table}[H]
  \caption{Model sampling configuration, as frozen in the run manifests.}
  \label{tab:api-config}
  \centering
  \setlength{\tabcolsep}{0pt}
  \begin{tabular}{@{}p{1.34in}@{\hspace{10pt}}%
      p{\dimexpr(\linewidth-1.34in-10pt)/3\relax}%
      p{\dimexpr(\linewidth-1.34in-10pt)/3\relax}%
      p{\dimexpr(\linewidth-1.34in-10pt)/3\relax}@{}}
    \toprule
    & GPT & Claude & Gemini\\
    \midrule
    Model identifier & gpt-5.6-sol & claude-opus-4-8 & gemini-3.5-flash\\
    API & Responses & Messages & GenerateContent\\
    Reasoning control & reasoning effort & adaptive thinking & thinking level\\
    Settings & low, medium & low, medium & low, medium\\
    Output token cap & 8{,}192 & 8{,}192 & 8{,}192\\
    Text verbosity & low & --- & ---\\
    Image detail & high & --- & ---\\
    Thinking display & --- & omitted & ---\\
    \midrule
    Sampling & \multicolumn{3}{l}{provider defaults; no temperature or
      top-p override}\\
    Prompt conditions & \multicolumn{3}{l}{plain, persona
      (Supplementary Fig.~\ref{fig:app-api-prompt})}\\
    Interaction & \multicolumn{3}{l}{single turn, no feedback; no tools; no
      streaming}\\
    Collection & \multicolumn{3}{l}{provider batch APIs}\\
    \bottomrule
  \end{tabular}
\end{table}

\begin{table}[H]
  \caption{Data underlying the analyses.  Human counts are retained
  participants and attempted responses; model counts are scored response
  rows over four conditions (two reasoning-effort settings $\times$ two prompts) at 100 samples
  per puzzle and condition.  These are collection totals; primary analyses
  use the 100 low-effort, plain-prompt samples for each model and puzzle.
  Rates are correct for humans and valid for models.}
  \label{tab:data}
  \centering
  \begin{tabular}{@{}lrrl@{}}
    \toprule
    Source & Responses & Correct/valid & Collected \\
    \midrule
    Human, main battery (104 participants)      & 10{,}400 & 90.5\%       & 2026-06 \\
    Human, modules (417 participants, 2 waves)  & 25{,}647 & 92.3\%       & 2026-06 \\
    GPT, main battery                       & 40{,}000 & 98.0--99.4\% & 2026-07 \\
    GPT, modules                            & 68{,}000 & 99.8--99.9\% & 2026-07 \\
    Claude, main battery                        & 40{,}000 & 99.6--99.8\% & 2026-06 \\
    Claude, modules                             & 68{,}000 & 99.3--99.4\% & 2026-06 \\
    Gemini, main battery                        & 40{,}000 & 98.9--99.2\% & 2026-07 \\
    Gemini, modules                             & 68{,}000 & 99.7--99.9\% & 2026-07 \\
    \bottomrule
  \end{tabular}
\end{table}
 \FloatBarrier

\section{Main study}\label{app:condition-results}
Unless stated otherwise, model estimates use the low-effort, plain-prompt
condition.

\subsection{Distributional similarity}\label{app:tv-analysis}

\subsubsection{Point estimates and uncertainty}\label{app:tv-estimation}
\label{supp:response-bootstrap}
Table~\ref{tab:tv-global} compares bootstrap intervals obtained by resampling
puzzles, trials, or both. \emph{Puzzles} uses the whole-puzzle bootstrap in Methods,
Section~\ref{supp:puzzle-bootstrap}. \emph{Trials} fixes the battery and
resamples the 104 retained Human participant profiles and the 100 original
API requests within each model--puzzle cell, reapplying the correct/valid
filter. \emph{Both} combines these schemes; repeated copies of a puzzle reuse
its trial-resampled distribution under multiplicity weights. Draws are shared
across source pairs, Uniform is fixed, and a replicate with an empty retained
cell is redrawn. The three intervals are not additive variance components.

\begin{table}[H]
  \color{black}
  \captionsetup{font+={color=black}}
  \caption{Pairwise mean total-variation distance across the 100-puzzle
  battery.  Uniform assigns equal mass to a puzzle's valid solution classes.}
  \label{tab:tv-global}
  \centering
  \setlength{\tabcolsep}{4pt}
  \begin{tabular}{@{}lrrrr@{}}
    \toprule
    & & \multicolumn{3}{c}{95\% bootstrap CI} \\
    \cmidrule(lr){3-5}
    Comparison & Mean TV & Puzzles & Trials & Both \\
    \midrule
    Uniform vs Human & $0.30$ & $[0.29,0.32]$ & $[0.30,0.33]$ & $[0.29,0.34]$ \\
    Uniform vs GPT & $0.65$ & $[0.63,0.68]$ & $[0.65,0.66]$ & $[0.63,0.68]$ \\
    Uniform vs Claude & $0.55$ & $[0.52,0.58]$ & $[0.54,0.56]$ & $[0.52,0.58]$ \\
    Uniform vs Gemini & $0.56$ & $[0.53,0.59]$ & $[0.56,0.57]$ & $[0.53,0.59]$ \\
    \midrule
    Human vs GPT & $0.54$ & $[0.51,0.57]$ & $[0.53,0.56]$ & $[0.51,0.58]$ \\
    Human vs Claude & $0.45$ & $[0.42,0.48]$ & $[0.44,0.47]$ & $[0.42,0.49]$ \\
    Human vs Gemini & $0.46$ & $[0.43,0.49]$ & $[0.45,0.49]$ & $[0.43,0.50]$ \\
    \midrule
    GPT vs Claude & $0.36$ & $[0.32,0.41]$ & $[0.36,0.38]$ & $[0.32,0.41]$ \\
    GPT vs Gemini & $0.38$ & $[0.33,0.43]$ & $[0.37,0.39]$ & $[0.33,0.43]$ \\
    Claude vs Gemini & $0.32$ & $[0.28,0.36]$ & $[0.32,0.34]$ & $[0.29,0.37]$ \\
    \bottomrule
  \end{tabular}
\end{table}
 
\subsubsection{Robustness to alternative distribution measures}
\label{app:metric-comparison}\label{app:five-source-variants}

\paragraph{Alternative distribution measures.}\label{app:robustness-measures}
\leavevmode
The finite symmetric distance comparisons include Jensen--Shannon and
Hellinger distances.  Let
$m=(p+q)/2$ and define
\begin{align}
  d_{\mathrm{JS}}(p,q)
  &=\left\{\frac12 D_{\mathrm{KL}}(p\Vert m)
      +\frac12 D_{\mathrm{KL}}(q\Vert m)\right\}^{1/2},
      \label{eq:js-distance}\\
  d_{\mathrm{H}}(p,q)
  &=\left\{\frac12\sum_{j\in\mathcal J_t}
      \bigl(\sqrt{p(j)}-\sqrt{q(j)}\bigr)^2\right\}^{1/2},
      \label{eq:hellinger-distance}
\end{align}
where $D_{\mathrm{KL}}(p\Vert q)=\sum_j p(j)\log_2[p(j)/q(j)]$ is the
Kullback--Leibler divergence.
These are the \mbox{Jensen--Shannon distance} and Hellinger distance, respectively;
both lie in $[0,1]$ and remain finite when empirical supports differ.

For strictly positive distributions with nonconstant log-density vectors, we
define the within-puzzle log-density correlation as
\begin{equation}
  \rho_{\log,t}(p,q)
  =\operatorname{corr}\!\left(
    \left\{\bigl(\log p(j),\log q(j)\bigr):j\in\mathcal J_t\right\}
  \right).
  \label{eq:log-density-correlation}
\end{equation}
The log base does not affect the correlation.  The statistic is undefined when
either log-density vector is constant, so comparisons with Uniform are not
reported. Log-density correlation cannot detect a pure concentration
difference: $\rho_{\log}(p,q)=1$ whenever $q_j\propto p_j^a$ with $a>0$.

For an empirical distribution $p$, let
$\mathcal A(p)=\arg\max_{j\in\mathcal J_t}p(j)$ retain every maximizer.  The
tie-aware modal agreement is
\begin{equation}
  M_t(p,q)
  =\mathbf 1\!\left\{\mathcal A(p)\cap\mathcal A(q)\ne\varnothing\right\}.
  \label{eq:modal-agreement}
\end{equation}
It equals one when the two empirical distributions share at least one
most-frequent solution and zero otherwise.  Uniform is omitted from modal
comparisons because every valid solution is one of its modes, which would make
its agreement identically one.

\paragraph{Half-count smoothing.}\label{app:log-density-smoothing}
For source $s$, condition $c$, and puzzle $t$, let $N_{tsc}(j)$ count retained
responses selecting solution $j$, with total $n_{tsc}$. For log-density correlation, we make log
densities finite by adding half a count to each of the $k_t$ valid solution classes:
\begin{equation}
  \widetilde p^{(1/2)}_{tsc}(j)
  =\frac{N_{tsc}(j)+1/2}{n_{tsc}+k_t/2}.
  \label{eq:half-count-density}
\end{equation}
In the four-source comparisons below, only log-density comparisons use this
smoothing; the other measures use empirical proportions. 
 
\paragraph{Common comparison design.}\label{app:metric-estimation}
The four-source comparison uses the six unordered pairs among Human,
GPT, Claude and Gemini; Uniform is excluded for the reasons given above.
Measures use the primary conditions and equal-family averaging in
Section~\ref{app:tv-estimation}, with smoothing as specified in
Section~\ref{app:log-density-smoothing}. A log-density correlation is omitted
only when a smoothed log-density vector is constant.

Table~\ref{tab:metric-comparison-family} reports family estimates and
equal-family means with pointwise 95\% puzzle-bootstrap intervals
(Section~\ref{supp:puzzle-bootstrap}). Draws are shared across measures and
source pairs; corresponding TV estimates are shown in Figure~\ref{fig:tv-source-geometry}.

\begin{table}[!htbp]
  \caption{Alternative distribution measures by puzzle family in the primary low-effort, plain-prompt condition; corresponding TV estimates are shown in Figure~\ref{fig:tv-source-geometry}. Each cell is the metric estimate followed by its pointwise 95\% puzzle-bootstrap interval; the Mean column weights the five family estimates equally.}
  \label{tab:metric-comparison-family}
  \centering
  \small
  \setlength{\tabcolsep}{0.8pt}
  \renewcommand{\arraystretch}{0.94}
  \begin{adjustbox}{max width=\linewidth}
\begin{tabular}{@{}lcccccc@{}}
    \toprule
    Comparison & Mean & Arithmetic & Maze & Rooks & Minesweeper & Sudoku \\
    \midrule
    \addlinespace[0.35em]
    \multicolumn{7}{@{}l}{\textit{Jensen--Shannon distance (lower is closer)}} \\
    Human vs GPT & $0.56[0.53,0.59]$ & $0.44[0.38,0.50]$ & $0.41[0.35,0.49]$ & $0.62[0.56,0.68]$ & $0.64[0.60,0.69]$ & $0.68[0.63,0.73]$ \\
    Human vs Claude & $0.46[0.44,0.49]$ & $0.41[0.36,0.45]$ & $0.40[0.35,0.46]$ & $0.55[0.50,0.61]$ & $0.44[0.39,0.49]$ & $0.49[0.43,0.55]$ \\
    Human vs Gemini & $0.48[0.45,0.51]$ & $0.51[0.45,0.56]$ & $0.31[0.26,0.36]$ & $0.55[0.49,0.61]$ & $0.54[0.46,0.61]$ & $0.48[0.41,0.54]$ \\
    GPT vs Claude & $0.41[0.37,0.45]$ & $0.44[0.38,0.50]$ & $0.38[0.30,0.46]$ & $0.36[0.27,0.44]$ & $0.53[0.44,0.62]$ & $0.35[0.28,0.43]$ \\
    GPT vs Gemini & $0.41[0.36,0.46]$ & $0.53[0.43,0.62]$ & $0.31[0.24,0.39]$ & $0.31[0.23,0.40]$ & $0.52[0.38,0.66]$ & $0.38[0.27,0.48]$ \\
    Claude vs Gemini & $0.36[0.33,0.39]$ & $0.49[0.43,0.54]$ & $0.29[0.22,0.35]$ & $0.30[0.23,0.38]$ & $0.35[0.28,0.44]$ & $0.37[0.29,0.45]$ \\
    \addlinespace[0.35em]
    \multicolumn{7}{@{}l}{\textit{Hellinger distance (lower is closer)}} \\
    Human vs GPT & $0.52[0.50,0.55]$ & $0.40[0.34,0.46]$ & $0.37[0.31,0.44]$ & $0.58[0.52,0.63]$ & $0.61[0.57,0.65]$ & $0.65[0.60,0.69]$ \\
    Human vs Claude & $0.41[0.39,0.44]$ & $0.36[0.32,0.40]$ & $0.37[0.31,0.42]$ & $0.50[0.44,0.55]$ & $0.40[0.34,0.45]$ & $0.44[0.38,0.50]$ \\
    Human vs Gemini & $0.44[0.41,0.46]$ & $0.45[0.40,0.50]$ & $0.28[0.23,0.33]$ & $0.50[0.44,0.55]$ & $0.51[0.43,0.58]$ & $0.44[0.38,0.50]$ \\
    GPT vs Claude & $0.38[0.35,0.41]$ & $0.39[0.34,0.45]$ & $0.34[0.26,0.41]$ & $0.33[0.25,0.41]$ & $0.51[0.42,0.59]$ & $0.34[0.27,0.42]$ \\
    GPT vs Gemini & $0.38[0.34,0.42]$ & $0.48[0.39,0.57]$ & $0.27[0.21,0.34]$ & $0.28[0.21,0.36]$ & $0.50[0.36,0.63]$ & $0.36[0.26,0.46]$ \\
    Claude vs Gemini & $0.32[0.30,0.35]$ & $0.44[0.38,0.49]$ & $0.26[0.20,0.31]$ & $0.27[0.21,0.34]$ & $0.32[0.25,0.40]$ & $0.33[0.26,0.40]$ \\
    \addlinespace[0.35em]
    \multicolumn{7}{@{}l}{\textit{Log-density correlation (higher is closer)}} \\
    Human vs GPT & $0.45[0.37,0.52]$ & $0.70[0.62,0.77]$ & $0.59[0.36,0.77]$ & $0.25[0.03,0.45]$ & $0.20[-0.02,0.41]$ & $0.50[0.35,0.65]$ \\
    Human vs Claude & $0.53[0.45,0.60]$ & $0.76[0.72,0.80]$ & $0.73[0.58,0.83]$ & $0.20[-0.02,0.40]$ & $0.42[0.17,0.66]$ & $0.55[0.39,0.70]$ \\
    Human vs Gemini & $0.53[0.45,0.60]$ & $0.55[0.44,0.67]$ & $0.78[0.67,0.86]$ & $0.21[-0.05,0.45]$ & $0.47[0.26,0.65]$ & $0.63[0.48,0.75]$ \\
    GPT vs Claude & $0.66[0.60,0.71]$ & $0.71[0.62,0.78]$ & $0.69[0.52,0.83]$ & $0.77[0.70,0.83]$ & $0.38[0.17,0.56]$ & $0.76[0.69,0.83]$ \\
    GPT vs Gemini & $0.66[0.60,0.72]$ & $0.61[0.49,0.74]$ & $0.76[0.64,0.86]$ & $0.83[0.77,0.89]$ & $0.38[0.13,0.61]$ & $0.73[0.63,0.82]$ \\
    Claude vs Gemini & $0.73[0.68,0.78]$ & $0.62[0.51,0.72]$ & $0.77[0.59,0.88]$ & $0.78[0.67,0.86]$ & $0.72[0.59,0.83]$ & $0.78[0.70,0.84]$ \\
    \addlinespace[0.35em]
    \multicolumn{7}{@{}l}{\textit{Tie-aware modal agreement (higher is closer)}} \\
    Human vs GPT & $0.43[0.33,0.52]$ & $0.45[0.25,0.70]$ & $0.60[0.40,0.80]$ & $0.30[0.10,0.50]$ & $0.35[0.15,0.55]$ & $0.45[0.25,0.65]$ \\
    Human vs Claude & $0.43[0.34,0.52]$ & $0.35[0.15,0.55]$ & $0.70[0.50,0.90]$ & $0.25[0.10,0.45]$ & $0.35[0.15,0.55]$ & $0.50[0.30,0.70]$ \\
    Human vs Gemini & $0.43[0.34,0.52]$ & $0.20[0.05,0.40]$ & $0.75[0.55,0.90]$ & $0.20[0.05,0.40]$ & $0.50[0.30,0.70]$ & $0.50[0.30,0.70]$ \\
    GPT vs Claude & $0.68[0.59,0.77]$ & $0.70[0.50,0.90]$ & $0.50[0.30,0.70]$ & $0.80[0.60,0.95]$ & $0.45[0.25,0.65]$ & $0.95[0.85,1.00]$ \\
    GPT vs Gemini & $0.54[0.45,0.63]$ & $0.35[0.15,0.55]$ & $0.50[0.30,0.70]$ & $0.75[0.55,0.95]$ & $0.35[0.15,0.55]$ & $0.75[0.55,0.90]$ \\
    Claude vs Gemini & $0.62[0.53,0.70]$ & $0.30[0.10,0.50]$ & $0.75[0.55,0.90]$ & $0.75[0.55,0.95]$ & $0.60[0.40,0.80]$ & $0.70[0.50,0.90]$ \\
    \bottomrule
  \end{tabular}
\end{adjustbox}
\end{table}
 
The overall conclusion is stable across metrics.  Under TV,
Jensen--Shannon, and Hellinger, every global model--model distance is smaller
than every Human--model distance.  The same separation appears when similarity
is measured instead: model--model log-density correlations are
$0.66$--$0.73$, compared with $0.45$--$0.53$ for Human--model pairs, and
model--model modal agreement is $0.54$--$0.68$, compared with $0.43$ for each
Human--model pair. Family-level exceptions occur, particularly in Arithmetic.

\FloatBarrier
\subsection{Entropy}\label{app:entropy-analysis}
\label{app:entropy-estimation}\label{app:entropy-profile-results}

Empirical entropy is downward-biased with finitely many responses; no
correction is applied. Puzzle-bootstrap intervals hold response distributions
fixed, so they neither include response-level uncertainty nor remove this bias.

Table~\ref{tab:entropy-svd} reports the family coefficients and source scores
for the entropy decomposition in Section~\ref{sec:results-entropy}.

\begin{table}[H]
  \caption{Centered entropy-profile SVD.\@  Each source is demeaned across
  puzzle families before decomposition; percentages are shares of the
  centered matrix's Frobenius energy.}
  \label{tab:entropy-svd}
  \centering
  \setlength{\tabcolsep}{10pt}
  \begin{adjustbox}{max width=\linewidth}
  \begin{tabular}[t]{@{}lrr@{}}
    \toprule
    \multicolumn{3}{@{}l}{\textit{Panel A: family coefficients}} \\
    \addlinespace[2pt]
    Puzzle family & SV1 (81.1\%) & SV2 (9.5\%) \\
    \midrule
    Arithmetic & $0.57$ & $-0.63$ \\
    Maze & $0.47$ & $0.74$ \\
    Rooks & $-0.14$ & $0.01$ \\
    Minesweeper & $-0.33$ & $-0.22$ \\
    Sudoku & $-0.57$ & $0.10$ \\
    \bottomrule
  \end{tabular}
  \hspace{0.07\linewidth}
  \begin{tabular}[t]{@{}lrr@{}}
    \toprule
    \multicolumn{3}{@{}l}{\textit{Panel B: source scores}} \\
    \addlinespace[2pt]
    Source & SV1 & SV2 \\
    \midrule
    Human & $-0.24$ & $-0.03$ \\
    GPT & $0.37$ & $0.03$ \\
    Claude & $0.03$ & $-0.15$ \\
    Gemini & $0.20$ & $-0.06$ \\
    \bottomrule
  \end{tabular}
  \end{adjustbox}
\end{table}
 
\subsubsection{Cross-puzzle entropy correlations}
\label{app:entropy-correlations}
Supplementary Figure~\ref{fig:entropy-correlation-pooled} compares normalized
entropy between sources over the 100-puzzle main battery. Pearson
correlations across these paired values are positive for all three
model--model pairs ($0.23$--$0.43$); Human correlations with GPT, Claude
and Gemini are $-0.33$, $0.01$ and $-0.14$, respectively. These pooled
correlations combine within-family and between-family variation.
{Uniform is omitted because its normalized entropy is one for
every puzzle, making its correlation with any source undefined.}

\begin{figure}[!t]
  \centering
  \includegraphics[width=\linewidth]{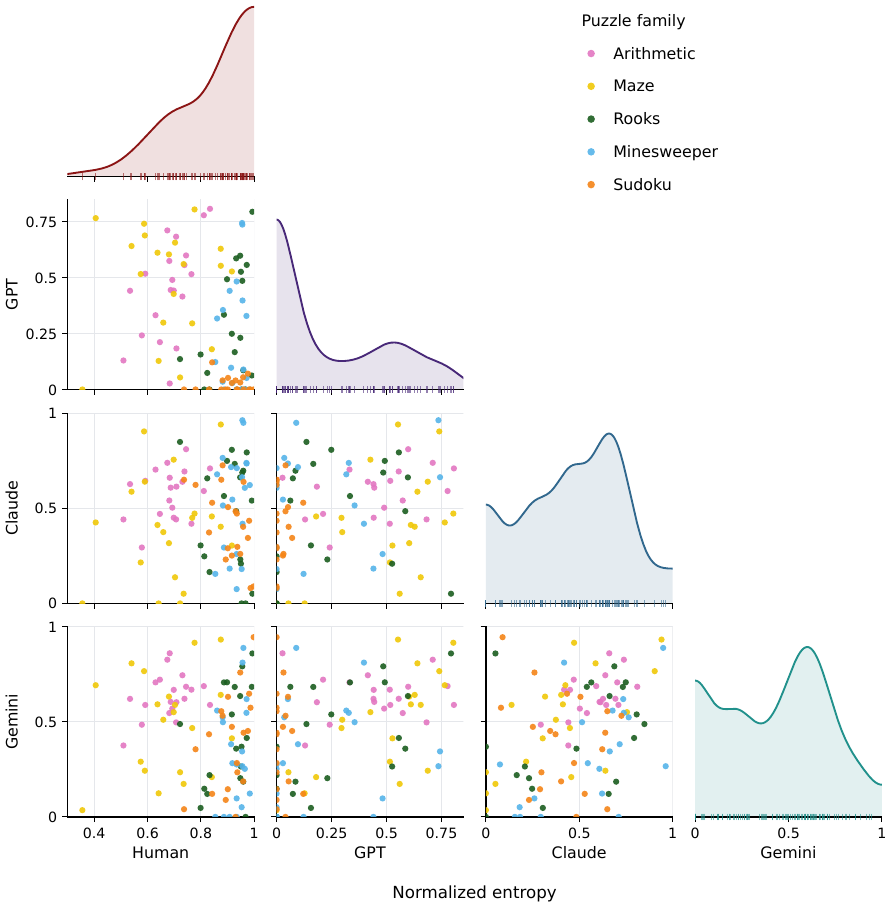}
  \caption{\textbf{Per-puzzle normalized-entropy relationships.}  Each panel
  compares one source pair over the 100-puzzle main battery. Diagonal panels
  show kernel density estimates of each source's normalized entropy
  (bandwidth 0.075, reflected at 0 and 1); ticks mark the individual puzzles.}
  \label{fig:entropy-correlation-pooled}
\end{figure}

\FloatBarrier

\clearpage
\subsection{Relative difficulty}\label{supp:effort-summaries}
\label{app:relative-difficulty}
Relative-difficulty scores are defined in main Section~\ref{sec:results-difficulty};
uncertainty follows Section~\ref{sec:methods-statistics}.

\begin{table}[!htbp]
  \caption{Pearson correlation of per-puzzle relative-difficulty scores
  by puzzle family.  The Mean column weights the five family correlations
  equally.  Brackets are 95\% paired-puzzle bootstrap intervals.}
  \label{tab:effort-correlation}
  \centering
  \small
  \setlength{\tabcolsep}{1.2pt}
  \renewcommand{\arraystretch}{1.08}
  \begin{tabular}{@{}lrrrrrr@{}}
    \toprule
    & \multicolumn{6}{c}{Pearson correlation [95\% CI]} \\
    \cmidrule(l){2-7}
    Source pair & Mean & Arithmetic & Maze & Rooks & Minesweeper & Sudoku \\
    \midrule
    Human vs GPT & $0.59[0.47,0.71]$ & $0.71[0.54,0.85]$ & $0.51[0.20,0.82]$ & $0.85[0.74,0.92]$ & $0.72[0.48,0.87]$ & $0.17[-0.32,0.58]$ \\
    Human vs Claude & $0.72[0.60,0.80]$ & $0.56[0.20,0.81]$ & $0.72[0.43,0.92]$ & $0.77[0.60,0.89]$ & $0.78[0.62,0.90]$ & $0.76[0.47,0.91]$ \\
    Human vs Gemini & $0.62[0.48,0.72]$ & $0.15[-0.37,0.52]$ & $0.74[0.50,0.89]$ & $0.86[0.76,0.94]$ & $0.73[0.55,0.86]$ & $0.62[0.29,0.84]$ \\
    GPT vs Claude & $0.71[0.56,0.81]$ & $0.68[0.50,0.85]$ & $0.87[0.67,0.96]$ & $0.90[0.81,0.97]$ & $0.75[0.46,0.93]$ & $0.32[-0.32,0.75]$ \\
    GPT vs Gemini & $0.60[0.44,0.74]$ & $0.27[-0.32,0.67]$ & $0.76[0.51,0.89]$ & $0.92[0.82,0.97]$ & $0.80[0.60,0.93]$ & $0.24[-0.28,0.69]$ \\
    Claude vs Gemini & $0.72[0.61,0.81]$ & $0.37[-0.09,0.72]$ & $0.80[0.59,0.90]$ & $0.87[0.68,0.96]$ & $0.90[0.80,0.97]$ & $0.68[0.44,0.83]$ \\
    \bottomrule
  \end{tabular}
\end{table}
 \FloatBarrier

\begin{figure}[H]
  \centering
  \includegraphics[width=0.62\linewidth]{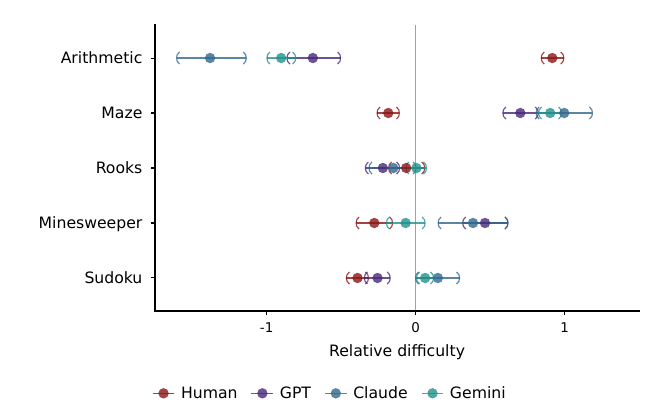}
  \caption{\textbf{Relative difficulty by puzzle family.} Points are
  family means. Bars show pointwise 95\%
  intervals from 5,000 whole-puzzle bootstrap resamples within family,
  with each source's mean recomputed in every resample.}
  \label{fig:effort-by-type}
\end{figure}
\FloatBarrier
 
\FloatBarrier
\clearpage
\subsection{Effects of reasoning-effort settings and prompting}\label{supp:condition-comparisons}
\label{app:tv-conditions}
\label{supp:condition-changes}

\begin{table}[H]
  \caption{Condition contrasts in model--Human mean TV.
  Estimates and 95\% paired-puzzle bootstrap intervals are multiplied
  by 100 (Section~\ref{supp:puzzle-bootstrap}). Positive values indicate
  greater distance from Human; Mean weights the five families equally.
  Bold estimates have intervals excluding zero. Intervals are pointwise
  and unadjusted for multiple comparisons.}
  \label{tab:tv-condition-contrasts}
  \centering
  \small
  \setlength{\tabcolsep}{3pt}
  \renewcommand{\arraystretch}{1.15}
  \begin{tabular*}{\linewidth}{@{\extracolsep{\fill}}lcccccc@{}}
    \toprule
    Contrast & Mean & Arithmetic & Maze & Rooks & Minesweeper & Sudoku \\
    \midrule
    \multicolumn{7}{@{}l}{\textbf{GPT}} \\
    \addlinespace[2pt]
    \shortstack[l]{Persona $-$ plain\\(low effort)} & \shortstack{$0.82$\\[1pt]{\scriptsize$[-0.35,1.95]$}} & \shortstack{$1.53$\\[1pt]{\scriptsize$[-1.51,4.67]$}} & \shortstack{$1.14$\\[1pt]{\scriptsize$[-1.98,4.35]$}} & \shortstack{$1.17$\\[1pt]{\scriptsize$[-0.95,3.49]$}} & \shortstack{$0.26$\\[1pt]{\scriptsize$[-2.43,2.99]$}} & \shortstack{$0.00$\\[1pt]{\scriptsize$[-0.50,0.50]$}} \\[4pt]
    \shortstack[l]{Persona $-$ plain\\(medium effort)} & \shortstack{$\mathbf{-1.08}$\\[1pt]{\scriptsize$[-1.98,-0.21]$}} & \shortstack{$-1.42$\\[1pt]{\scriptsize$[-4.03,1.11]$}} & \shortstack{$0.03$\\[1pt]{\scriptsize$[-2.27,2.19]$}} & \shortstack{$\mathbf{-2.55}$\\[1pt]{\scriptsize$[-4.90,-0.57]$}} & \shortstack{$\mathbf{-1.86}$\\[1pt]{\scriptsize$[-4.15,-0.14]$}} & \shortstack{$0.40$\\[1pt]{\scriptsize$[0.00,0.85]$}} \\[4pt]
    \shortstack[l]{Medium $-$ low\\(plain prompt)} & \shortstack{$1.11$\\[1pt]{\scriptsize$[-0.07,2.30]$}} & \shortstack{$1.95$\\[1pt]{\scriptsize$[-1.18,5.13]$}} & \shortstack{$-0.27$\\[1pt]{\scriptsize$[-3.97,2.99]$}} & \shortstack{$\mathbf{2.91}$\\[1pt]{\scriptsize$[0.75,5.43]$}} & \shortstack{$1.04$\\[1pt]{\scriptsize$[-1.99,3.92]$}} & \shortstack{$-0.05$\\[1pt]{\scriptsize$[-0.60,0.50]$}} \\[4pt]
    \shortstack[l]{Medium $-$ low\\(persona prompt)} & \shortstack{$-0.78$\\[1pt]{\scriptsize$[-2.08,0.52]$}} & \shortstack{$-1.00$\\[1pt]{\scriptsize$[-3.77,1.69]$}} & \shortstack{$-1.38$\\[1pt]{\scriptsize$[-5.76,2.88]$}} & \shortstack{$-0.81$\\[1pt]{\scriptsize$[-3.57,1.89]$}} & \shortstack{$-1.09$\\[1pt]{\scriptsize$[-4.12,1.47]$}} & \shortstack{$\mathbf{0.35}$\\[1pt]{\scriptsize$[0.10,0.70]$}} \\[4pt]
    \midrule
    \multicolumn{7}{@{}l}{\textbf{Claude}} \\
    \addlinespace[2pt]
    \shortstack[l]{Persona $-$ plain\\(low effort)} & \shortstack{$-0.26$\\[1pt]{\scriptsize$[-1.29,0.78]$}} & \shortstack{$0.74$\\[1pt]{\scriptsize$[-1.40,3.02]$}} & \shortstack{$\mathbf{-2.45}$\\[1pt]{\scriptsize$[-4.89,-0.25]$}} & \shortstack{$2.00$\\[1pt]{\scriptsize$[-0.24,4.40]$}} & \shortstack{$-0.52$\\[1pt]{\scriptsize$[-2.37,1.34]$}} & \shortstack{$-1.08$\\[1pt]{\scriptsize$[-3.74,1.54]$}} \\[4pt]
    \shortstack[l]{Persona $-$ plain\\(medium effort)} & \shortstack{$-1.00$\\[1pt]{\scriptsize$[-2.09,0.11]$}} & \shortstack{$0.54$\\[1pt]{\scriptsize$[-1.60,2.67]$}} & \shortstack{$2.11$\\[1pt]{\scriptsize$[-0.36,4.90]$}} & \shortstack{$-2.20$\\[1pt]{\scriptsize$[-4.93,0.51]$}} & \shortstack{$0.27$\\[1pt]{\scriptsize$[-2.03,2.64]$}} & \shortstack{$\mathbf{-5.72}$\\[1pt]{\scriptsize$[-8.23,-3.20]$}} \\[4pt]
    \shortstack[l]{Medium $-$ low\\(plain prompt)} & \shortstack{$\mathbf{1.85}$\\[1pt]{\scriptsize$[0.66,3.03]$}} & \shortstack{$-1.37$\\[1pt]{\scriptsize$[-3.73,1.33]$}} & \shortstack{$\mathbf{-2.57}$\\[1pt]{\scriptsize$[-5.56,-0.08]$}} & \shortstack{$2.06$\\[1pt]{\scriptsize$[-1.00,4.99]$}} & \shortstack{$\mathbf{2.72}$\\[1pt]{\scriptsize$[0.07,5.42]$}} & \shortstack{$\mathbf{8.42}$\\[1pt]{\scriptsize$[5.95,10.82]$}} \\[4pt]
    \shortstack[l]{Medium $-$ low\\(persona prompt)} & \shortstack{$1.11$\\[1pt]{\scriptsize$[-0.27,2.48]$}} & \shortstack{$-1.57$\\[1pt]{\scriptsize$[-4.91,1.72]$}} & \shortstack{$1.99$\\[1pt]{\scriptsize$[-1.17,5.29]$}} & \shortstack{$-2.14$\\[1pt]{\scriptsize$[-6.11,1.44]$}} & \shortstack{$\mathbf{3.51}$\\[1pt]{\scriptsize$[1.15,5.90]$}} & \shortstack{$\mathbf{3.78}$\\[1pt]{\scriptsize$[1.19,6.36]$}} \\[4pt]
    \midrule
    \multicolumn{7}{@{}l}{\textbf{Gemini}} \\
    \addlinespace[2pt]
    \shortstack[l]{Persona $-$ plain\\(low effort)} & \shortstack{$0.87$\\[1pt]{\scriptsize$[-0.43,2.20]$}} & \shortstack{$-0.36$\\[1pt]{\scriptsize$[-2.95,2.15]$}} & \shortstack{$-1.55$\\[1pt]{\scriptsize$[-4.18,0.87]$}} & \shortstack{$1.00$\\[1pt]{\scriptsize$[-1.39,3.60]$}} & \shortstack{$\mathbf{2.65}$\\[1pt]{\scriptsize$[0.36,5.31]$}} & \shortstack{$2.63$\\[1pt]{\scriptsize$[-1.33,6.73]$}} \\[4pt]
    \shortstack[l]{Persona $-$ plain\\(medium effort)} & \shortstack{$0.89$\\[1pt]{\scriptsize$[-0.47,2.26]$}} & \shortstack{$0.17$\\[1pt]{\scriptsize$[-2.01,2.26]$}} & \shortstack{$0.06$\\[1pt]{\scriptsize$[-2.26,2.32]$}} & \shortstack{$-0.15$\\[1pt]{\scriptsize$[-2.83,2.68]$}} & \shortstack{$0.02$\\[1pt]{\scriptsize$[-4.53,3.92]$}} & \shortstack{$\mathbf{4.37}$\\[1pt]{\scriptsize$[1.20,7.85]$}} \\[4pt]
    \shortstack[l]{Medium $-$ low\\(plain prompt)} & \shortstack{$0.52$\\[1pt]{\scriptsize$[-1.10,2.11]$}} & \shortstack{$0.21$\\[1pt]{\scriptsize$[-4.45,4.45]$}} & \shortstack{$\mathbf{3.75}$\\[1pt]{\scriptsize$[1.00,6.37]$}} & \shortstack{$0.10$\\[1pt]{\scriptsize$[-3.12,3.20]$}} & \shortstack{$-0.61$\\[1pt]{\scriptsize$[-5.44,4.14]$}} & \shortstack{$-0.86$\\[1pt]{\scriptsize$[-3.53,1.73]$}} \\[4pt]
    \shortstack[l]{Medium $-$ low\\(persona prompt)} & \shortstack{$0.54$\\[1pt]{\scriptsize$[-1.13,2.16]$}} & \shortstack{$0.74$\\[1pt]{\scriptsize$[-3.25,4.69]$}} & \shortstack{$\mathbf{5.35}$\\[1pt]{\scriptsize$[2.06,8.78]$}} & \shortstack{$-1.05$\\[1pt]{\scriptsize$[-3.76,1.77]$}} & \shortstack{$-3.23$\\[1pt]{\scriptsize$[-7.64,0.09]$}} & \shortstack{$0.88$\\[1pt]{\scriptsize$[-3.19,4.78]$}} \\[4pt]
    \bottomrule
  \end{tabular*}
\end{table}
 \FloatBarrier
 \FloatBarrier
\section{Predictive value of solution features}\label{sec:results-features}
\label{app:features}

We fit two complementary regressions within each family.  The first predicts
the selected solution from candidate attributes; the second asks whether the
attributes of a selected solution distinguish each model from Human on a
held-out puzzle.

Table~\ref{tab:feature-accuracy-overview} summarizes normalized held-out
performance; family tables report coefficients and raw selected-solution
accuracy (Acc.) and balanced source-classification accuracy (BA).
Brackets give pointwise 95\% intervals; bold estimates have intervals
excluding zero (Section~\ref{app:feature-bootstrap}).
\subsection{Feature-based prediction}\label{sec:methods-prediction}
\label{app:feature-bootstrap}
\paragraph{Feature scaling.}

\label{app:retained-feature-definitions}

For any nonbinary numeric attribute $f_{tj}$ that is scaled before fitting, let
$f_t^{\min}=\min_{\ell\in\mathcal J_t}f_{t\ell}$ and
$f_t^{\max}=\max_{\ell\in\mathcal J_t}f_{t\ell}$.
For a coordinate that varies within the puzzle, the scaled value is
\begin{equation}
  \widetilde f_{tj}
    =\frac{f_{tj}-f_t^{\min}}{f_t^{\max}-f_t^{\min}},
  \qquad f_t^{\max}>f_t^{\min}.
  \label{eq:feature-within-puzzle-scaling}
\end{equation}
For Model/Human prediction, a feature that is constant within a puzzle takes
its mean scaled value over nonconstant training puzzles, weighting puzzles
equally and their valid solutions uniformly. The mean is re-estimated within
each bootstrap sample and validation fold, with a fallback of $1/2$ when no
training puzzle varies in that feature.
For selected-solution prediction, constant coordinates can be coded as zero
because a common shift of all candidate scores cancels in the conditional
probabilities. Varying binary coordinates retain their natural zero--one
values; constant binary coordinates follow the same replacement rule.
\paragraph{Regression models.}
\label{app:feature-models}

For solution class $j\in\mathcal{J}_t$ in puzzle $t$, let
$\boldsymbol{x}_{tj}$ denote its feature vector.
For each source $s$, a conditional multinomial model predicts the selected
solution:
\begin{equation}
  \widehat p^{\mathsf{choice}}_s(j\mid t)
  =\frac{\exp\!\left(\langle\boldsymbol{\theta}_s,\boldsymbol{x}_{tj}\rangle\right)}
    {\sum_{\ell\in\mathcal{J}_t}
      \exp\!\left(\langle\boldsymbol{\theta}_s,\boldsymbol{x}_{t\ell}\rangle\right)}.
  \label{eq:selected-solution-choice}
\end{equation}
For each model $m$, a separate binary logistic regression predicts whether a
response came from that model or a human:
\begin{equation}
  \widehat p^{\mathsf{source}}_m(m\mid t,j)
  =\frac{\exp\!\left(\alpha_m+
    \langle\boldsymbol{\beta}_m,\boldsymbol{x}_{tj}\rangle\right)}
  {1+\exp\!\left(\alpha_m+
    \langle\boldsymbol{\beta}_m,\boldsymbol{x}_{tj}\rangle\right)}.
  \label{eq:source-binary}
\end{equation}

\paragraph{Fitting.}

All regression puzzles come from the 100-puzzle main battery.  Each family
uses its 20 puzzles and is fit separately.  Primary-condition fits use only the
low-effort, plain-prompt cell.
Puzzles receive equal total weight in both prediction tasks.  In each binary
provider-versus-Human fit, the two source classes additionally receive equal
weight within puzzle. Features enter jointly without interactions.

\label{supp:optimizer-settings}
We fit the models by minimizing the weighted mean negative log likelihood
plus the L2 penalty $\tfrac{\lambda}{2}\|\boldsymbol{\theta}_s\|_2^2$ or
$\tfrac{\lambda}{2}\|\boldsymbol{\beta}_m\|_2^2$ on the feature coefficients,
with $\lambda=10^{-4}$; the intercept $\alpha_m$ is not penalized. We use
damped Newton iterations, using Armijo step halving with parameter $10^{-4}$.
The stopping criterion is a maximum absolute gradient of $10^{-8}$ for the
penalized objective. Singular Hessians use a pseudoinverse; a non-descent
direction is replaced by negative gradient descent. A safety limit of 10,000
iterations retains the current finite estimate.

\paragraph{Out-of-puzzle validation.}
\label{app:feature-validation}

Validation holds out one entire puzzle at a time. For selected-solution
prediction, let $A_{ts}$ contain all candidates with the largest held-out
fitted probability. For $B\subseteq\mathcal{J}_t$, write
$p_{ts}(B)=\sum_{j\in B}p_{ts}(j)$.  The puzzle's tie-aware top-choice accuracy
is $a_{ts}=p_{ts}(A_{ts})/|A_{ts}|$: a response in a $q$-way fitted tie
receives $1/q$ credit, while a unique maximizer gives ordinary accuracy.

For Model/Human prediction, let $D_{tm}$ contain the solution classes with
held-out fitted model probability at least $1/2$, assigning ties to the model,
and let $D_{tm}^{c}$ be its complement in $\mathcal J_t$.
Balanced accuracy on puzzle $t$ is
$b_{tm}=\tfrac12\{p_{tm}(D_{tm})+
p_{t,\mathsf{human}}(D_{tm}^{c})\}$.

Acc.\ averages $a_{ts}$ over the 20 held-out puzzles, and BA averages $b_{tm}$.

\paragraph{Regression uncertainty.}
Regression intervals use 2,000 whole-puzzle bootstrap draws. Each draw samples
20 puzzles with replacement, refits the coefficient model and reruns complete
leave-one-puzzle-out validation. All copies of a held-out puzzle are removed
from training, and its validation score is weighted by its multiplicity in
the draw. The same fixed draws are used for coefficients and validation,
with every returned estimate retained.
\paragraph{Prediction scores and normalization.}
\label{app:prediction-metrics}

For selected-solution prediction, random guessing gives accuracy $a_t^0=1/k_t$.
The attainable top-choice accuracy for the empirical solution distribution is
$a_{ts}^{\star}=\max_{j\in\mathcal J_t}p_{ts}(j)$.
For Model/Human prediction, the exact Bayes ceiling when solution identity
is observed is
\begin{equation}
  b_{tm}^{\mathrm{TV}}
    =\frac{1+\TVm(p_{tm},p_{t,\mathsf{human}})}{2}.
  \label{eq:source-tv-bound}
\end{equation}
Here $1/2$ is the balanced accuracy under random guessing.

For each task, accuracy is normalized within each held-out puzzle using
the accuracy under random guessing and its attainable ceiling, then averaged over the
puzzles in family $h$:
\begin{equation}
  \operatorname{Norm}^{\mathsf{choice}}_s
  =\frac{1}{|\mathcal T_h|}\sum_{t\in\mathcal T_h}
    \frac{a_{ts}-a_t^0}{a_{ts}^{\star}-a_t^0},
  \qquad
  \operatorname{Norm}^{\mathsf{source}}_m
  =\frac{1}{|\mathcal T_h|}\sum_{t\in\mathcal T_h}
    \frac{b_{tm}-1/2}{b_{tm}^{\mathrm{TV}}-1/2}.
  \label{eq:normalized-regression-accuracy}
\end{equation}
A summand is set to zero when its denominator is zero. Normalized scores can
be negative.
\begin{table}[!htbp]
\centering
\caption{\textbf{Normalized held-out prediction accuracy.} Accuracy is normalized within each held-out puzzle before averaging over 20 puzzles per family (Section~\ref{sec:methods-prediction}). Brackets give 95\% intervals from 2,000 whole-puzzle bootstrap refits. Model responses use low effort and the plain prompt.}
\label{tab:feature-accuracy-overview}
\small
\setlength{\tabcolsep}{5pt}
\renewcommand{\arraystretch}{1.15}
\begin{tabular}{@{}lcccc@{}}
\toprule
\multicolumn{5}{@{}l}{\textbf{a}\quad Selected-solution prediction} \\
Family & Human & GPT & Claude & Gemini \\
\midrule
Arithmetic & $\boldsymbol{0.79}[0.62,0.92]$ & $0.20[-0.10,0.57]$ & $0.18[-0.03,0.42]$ & $0.34[-0.01,0.61]$ \\
Maze & $\boldsymbol{0.67}[0.49,0.88]$ & $\boldsymbol{0.57}[0.30,0.78]$ & $\boldsymbol{0.61}[0.39,0.91]$ & $\boldsymbol{0.65}[0.41,0.90]$ \\
Rooks & $0.20[-0.17,0.49]$ & $-0.03[-0.16,0.44]$ & $0.25[-0.05,0.54]$ & $0.01[-0.15,0.41]$ \\
Minesweeper & $0.33[-0.31,0.57]$ & $0.01[-0.37,0.34]$ & $0.25[-0.23,0.46]$ & $0.24[-0.27,0.47]$ \\
Sudoku & $0.26[-0.09,0.49]$ & $0.10[-0.25,0.29]$ & $-0.21[-0.27,0.31]$ & $0.06[-0.33,0.28]$ \\
\midrule
\multicolumn{5}{@{}l}{\textbf{b}\quad Model-versus-Human prediction} \\
Family & Human & GPT & Claude & Gemini \\
\midrule
Arithmetic & --- & $0.10[-0.24,0.40]$ & $0.19[-0.15,0.48]$ & $\boldsymbol{0.52}[0.31,0.74]$ \\
Maze & --- & $-0.07[-0.44,0.48]$ & $0.38[-0.09,0.66]$ & $\boldsymbol{0.41}[0.02,0.66]$ \\
Rooks & --- & $0.31[-0.02,0.58]$ & $\boldsymbol{0.52}[0.21,0.72]$ & $\boldsymbol{0.37}[0.13,0.66]$ \\
Minesweeper & --- & $0.20[-0.39,0.45]$ & $0.20[-0.34,0.41]$ & $0.14[-0.18,0.44]$ \\
Sudoku & --- & $0.00[-0.48,0.32]$ & $-0.02[-0.52,0.26]$ & $0.16[-0.39,0.40]$ \\
\bottomrule
\end{tabular}
\end{table}
 
\FloatBarrier

\subsection{Maze}\label{sec:maze-geometric-features}
\label{app:maze-features}

Every valid path in maze $t$ has the common shortest length $L_t$.
For candidate $j\in\mathcal J_t$, write
$m_{tj1},\ldots,m_{tjL_t}$ for its move sequence and $E_{tj}$ for its set of
directed edges.

\Needspace{8\baselineskip}
\paragraph{Turn count.}
The raw turn count is
\begin{equation}
  \tau_{tj}
  =\sum_{r=2}^{L_t}
    \mathbf 1\{m_{tjr}\ne m_{tj,r-1}\}.
  \label{eq:maze-turn-count}
\end{equation}
It counts changes between consecutive move directions.

\paragraph{Mean background edge load.}
For each path, we average the fraction of other valid paths that use the same
edge:
\begin{equation}
  O_{tj}
  =\frac{1}{L_t(k_t-1)}
    \sum_{\substack{\ell\in\mathcal J_t\\ \ell\ne j}}
      |E_{tj}\cap E_{t\ell}|.
  \label{eq:maze-background-edge-load}
\end{equation}
Larger values indicate greater overlap with other valid paths.

\paragraph{Mean relative distance.}
At each step, including the start and goal, we measure the Euclidean
distance from each path's current position to the goal. We subtract the mean
distance across all valid paths at that same step, then average these centered
distances over the steps of each path. Lower values indicate paths that remain closer to the goal relative to the other paths at corresponding steps.
\begin{table}[H]
\caption{Selected-solution Maze regressions.}
\label{tab:maze-choice-results}
\label{tab:maze-extended-choice}
\centering
\regressiontableformat
\begin{tabular}{@{}lrrrr@{}}
\toprule
Source & $\widehat{\theta}_{\mathrm{turn}}$ & $\widehat{\theta}_{\mathrm{edge}}$ & $\widehat{\theta}_{\mathrm{rel}}$ & Acc. \\
\midrule
Human & $\boldsymbol{-1.23}[-1.84,-0.76]$ & $\boldsymbol{-1.19}[-1.74,-0.58]$ & $-0.32[-1.01,0.30]$ & $0.41[0.32,0.52]$ \\
GPT & $\boldsymbol{-1.65}[-3.12,-0.61]$ & $\boldsymbol{-1.66}[-2.60,-1.00]$ & $\boldsymbol{-0.98}[-2.23,-0.04]$ & $0.46[0.32,0.60]$ \\
Claude & $\boldsymbol{-2.47}[-3.93,-1.51]$ & $\boldsymbol{-2.01}[-3.16,-1.05]$ & $-0.84[-1.87,0.13]$ & $0.54[0.39,0.75]$ \\
Gemini & $\boldsymbol{-1.60}[-2.97,-0.61]$ & $\boldsymbol{-1.81}[-2.56,-1.01]$ & $0.08[-0.99,0.90]$ & $0.54[0.40,0.71]$ \\
\bottomrule
\end{tabular}
\end{table}

\begin{table}[H]
\caption{Source classification for Maze.}
\label{tab:maze-source-binary-results}
\label{tab:maze-extended-source}
\centering
\regressiontableformat
\begin{tabular}{@{}lrrrr@{}}
\toprule
Model/Human & $\widehat{\beta}_{\mathrm{turn}}$ & $\widehat{\beta}_{\mathrm{edge}}$ & $\widehat{\beta}_{\mathrm{rel}}$ & BA \\
\midrule
GPT & $-0.55[-1.44,0.55]$ & $-0.41[-1.18,0.11]$ & $-0.71[-1.52,0.28]$ & $0.46[0.39,0.58]$ \\
Claude & $\boldsymbol{-1.08}[-2.19,-0.25]$ & $-0.61[-1.44,0.11]$ & $-0.58[-1.44,0.44]$ & $0.57[0.47,0.63]$ \\
Gemini & $-0.25[-1.14,0.52]$ & $\boldsymbol{-0.46}[-0.99,-0.03]$ & $0.27[-0.42,1.07]$ & $0.55[0.50,0.60]$ \\
\bottomrule
\end{tabular}
\end{table}
 
\subsection{Arithmetic}\label{supp:feature-results-arithmetic}
\label{app:arithmetic-features}
For Arithmetic puzzle $t$, let $\mathcal E_t$ contain every admissible binary
parse tree that uses the four displayed token occurrences exactly once and
evaluates to 24. The class map $\kappa_t$ assigns each tree its canonical
answer class. For candidate $j$, its complete set of representatives is
\begin{equation}
  \mathcal R_t(j)
  =\{r\in\mathcal E_t:\kappa_t(r)=j\}.
  \label{eq:arithmetic-representative-set}
\end{equation}

Features aggregate over the complete representative set $\mathcal R_t(j)$.
Using a minimum encodes the availability of a low-burden realization, not a
claim that the respondent used the minimizing tree.

\paragraph{Minimum division count.}
Let $\operatorname{Int}(r)$ be the internal nodes of tree $r$, and let
$\operatorname{op}_r(v)$ be the operation at node $v$.
Define
\begin{equation}
  D_{tj}
  =\min_{r\in\mathcal R_t(j)}
    \sum_{v\in\operatorname{Int}(r)}
      \mathbf 1\{\operatorname{op}_r(v)=\div\}.
  \label{eq:arithmetic-min-division}
\end{equation}
\paragraph{Balanced-plan affordance.}
Let $\lambda(r)$ be the number of leaves under the root's left child.
A four-leaf tree has a balanced root split exactly when $\lambda(r)=2$.
Define
\begin{equation}
  H_{tj}
  =\mathbf 1\{\exists r\in\mathcal R_t(j):\lambda(r)=2\}.
  \label{eq:arithmetic-balanced-affordance}
\end{equation}
Thus $H_{tj}=1$ when at least one representative combines two two-number
subexpressions at the root.

\Needspace{8\baselineskip}
\paragraph{Minimum display-order inversions.}
Number the displayed token positions $1,\ldots,4$ and read a tree's leaves
from left to right.
Let $\Pi_t(r)$ be the set of value-preserving bijections from those leaves to
the displayed positions, and let $\pi_u$ be the position assigned to leaf
$u$.
Define
\begin{align}
  I_t(r)
    &=\min_{\boldsymbol{\pi}\in\Pi_t(r)}
      \sum_{1\le u<v\le4}\mathbf 1\{\pi_u>\pi_v\},\\*
  I_{tj}
    &=\min_{r\in\mathcal R_t(j)}I_t(r).
  \label{eq:arithmetic-min-display-inversions}
\end{align}
The first minimum prevents repeated equal numbers from receiving arbitrary
identities.
The second gives the least reordering required by any representative of the
class.
Because this is a presentation feature, it is recomputed for each displayed
number order.

\FloatBarrier
\begin{table}[H]
  \caption{Selected-solution Arithmetic regressions.}
  \label{tab:arithmetic-choice-results}
  \centering
  \regressiontableformat
  \begin{tabular}{@{}lrrrr@{}}
    \toprule
    Source & $\widehat{\theta}_{\mathrm{div}}$ & $\widehat{\theta}_{\mathrm{bal}}$ & $\widehat{\theta}_{\mathrm{inv}}$ & Acc. \\
    \midrule
    Human & $\boldsymbol{-0.64}[-0.89,-0.34]$ & $\boldsymbol{1.72}[1.49,1.96]$ & $\boldsymbol{-0.94}[-1.37,-0.58]$ & $0.40[0.33,0.46]$ \\
    GPT & $-0.62[-1.54,0.05]$ & $\boldsymbol{1.41}[0.90,2.08]$ & $-0.21[-1.12,0.69]$ & $0.23[0.08,0.47]$ \\
    Claude & $-0.08[-0.82,0.41]$ & $\boldsymbol{1.26}[0.68,1.91]$ & $-0.49[-1.09,0.09]$ & $0.21[0.13,0.31]$ \\
    Gemini & $0.14[-0.62,0.76]$ & $\boldsymbol{0.81}[0.44,1.17]$ & $\boldsymbol{0.60}[0.04,1.19]$ & $0.28[0.13,0.42]$ \\
    \bottomrule
  \end{tabular}
\end{table}

\begin{table}[H]
  \caption{Source classification for Arithmetic.}
  \label{tab:arithmetic-source-binary-results}
  \centering
  \regressiontableformat
  \begin{tabular}{@{}lrrrr@{}}
    \toprule
    Model/Human & $\widehat{\beta}_{\mathrm{div}}$ & $\widehat{\beta}_{\mathrm{bal}}$ & $\widehat{\beta}_{\mathrm{inv}}$ & BA \\
    \midrule
    GPT & $0.07[-0.83,0.69]$ & $-0.32[-0.70,0.14]$ & $\boldsymbol{0.66}[0.05,1.31]$ & $0.55[0.46,0.61]$ \\
    Claude & $0.59[-0.14,1.12]$ & $-0.48[-0.97,0.09]$ & $\boldsymbol{0.49}[0.04,1.03]$ & $0.55[0.48,0.61]$ \\
    Gemini & $\boldsymbol{0.90}[0.18,1.47]$ & $\boldsymbol{-0.96}[-1.43,-0.53]$ & $\boldsymbol{1.49}[0.87,2.21]$ & $0.65[0.60,0.70]$ \\
    \bottomrule
  \end{tabular}
\end{table}
 
\subsection{Rooks}\label{supp:feature-results-rooks}
\label{app:grid-features}

Let $n_t$ be the board side length, let $m_t=n_t-1$ be the required number of
rooks, and let $F_t\subseteq[n_t]\times[n_t]$ be the forbidden cells.
Candidate $j$ is a set
$P_{tj}\subseteq([n_t]\times[n_t])\setminus F_t$ of $m_t$ cells with no
repeated row or column.

\paragraph{Mean marginal cell rarity.}
For a free cell $x$, define its prevalence in the valid menu and the
candidate-level aggregate by
\begin{align}
  q_t(x)
    &=\frac{1}{k_t}\sum_{\ell\in\mathcal J_t}
      \mathbf 1\{x\in P_{t\ell}\},\\
  R_{tj}
    &=-\frac{1}{m_t}\sum_{x\in P_{tj}}\log q_t(x).
  \label{eq:grid-cell-rarity}
\end{align}
A high value means that the placement uses cells appearing in relatively few
valid answers.

\paragraph{Symmetric monotonicity.}
Order the occupied cells by displayed row as
$(r_{tj1},c_{tj1}),\allowbreak\ldots,\allowbreak(r_{tjm_t},c_{tjm_t})$.
Because columns do not repeat, define
\begin{align}
  I_{tj}
    &=\sum_{1\le a<b\le m_t}
      \mathbf 1\{c_{tja}>c_{tjb}\},
  & I_t^{\max}&=\binom{m_t}{2},\\
  M_{tj}
    &=\frac{\min\{I_{tj},I_t^{\max}-I_{tj}\}}{I_t^{\max}}.
  \label{eq:grid-symmetric-monotonicity}
\end{align}
Ascending and descending diagonal-like sequences both receive zero.
The feature rises as a placement departs from both monotone directions.
It is invariant to horizontal or vertical reflection but can change under an
arbitrary row or column permutation, so it is computed in the displayed
orientation.

\paragraph{Column-sequence direction changes.}
For $a=1,\ldots,m_t-1$, let
$s_{tja}=\operatorname{sgn}(c_{tj,a+1}-c_{tja})$.
Define
\begin{equation}
  Z_{tj}
  =\frac{1}{m_t-2}
    \sum_{a=2}^{m_t-1}
      \mathbf 1\{s_{tja}\ne s_{tj,a-1}\}.
  \label{eq:grid-direction-changes}
\end{equation}
This measures displayed zigzag complexity.
For the $4\times4$ puzzles, $M_{tj}=Z_{tj}/3$ and the two normalized
coordinates coincide whenever they vary.
The $5\times5$ and $6\times6$ puzzles break this dependence, making the
full-battery design matrix full rank.
\begin{table}[H]
  \caption{Selected-solution Rooks regressions.}
  \label{tab:grid-choice-results}
  \centering
  \regressiontableformat
  \begin{tabular}{@{}lrrrr@{}}
    \toprule
    Source & $\widehat{\theta}_{\mathrm{rarity}}$ & $\widehat{\theta}_{\mathrm{mono}}$ & $\widehat{\theta}_{\mathrm{change}}$ & Acc. \\
    \midrule
    Human & $0.05[-0.21,0.33]$ & $-0.33[-0.83,0.07]$ & $-0.28[-0.72,0.22]$ & $0.22[0.15,0.29]$ \\
    GPT & $\boldsymbol{-1.78}[-3.33,-0.94]$ & $0.15[-1.29,1.38]$ & $-0.24[-1.63,1.25]$ & $0.16[0.08,0.47]$ \\
    Claude & $\boldsymbol{-2.31}[-3.21,-1.70]$ & $-0.29[-1.36,0.70]$ & $-0.01[-0.93,0.84]$ & $0.33[0.17,0.51]$ \\
    Gemini & $\boldsymbol{-1.86}[-2.91,-1.29]$ & $0.26[-0.92,1.39]$ & $-0.15[-1.28,0.88]$ & $0.21[0.13,0.42]$ \\
    \bottomrule
  \end{tabular}
\end{table}

\begin{table}[H]
  \caption{Source classification for Rooks.}
  \label{tab:grid-source-binary-results}
  \centering
  \regressiontableformat
  \begin{tabular}{@{}lrrrr@{}}
    \toprule
    Model/Human & $\widehat{\beta}_{\mathrm{rarity}}$ & $\widehat{\beta}_{\mathrm{mono}}$ & $\widehat{\beta}_{\mathrm{change}}$ & BA \\
    \midrule
    GPT & $\boldsymbol{-1.67}[-2.77,-0.93]$ & $0.32[-0.75,1.18]$ & $-0.05[-0.89,0.83]$ & $0.61[0.50,0.70]$ \\
    Claude & $\boldsymbol{-2.10}[-2.71,-1.63]$ & $0.02[-0.79,0.79]$ & $0.19[-0.41,0.72]$ & $0.65[0.56,0.72]$ \\
    Gemini & $\boldsymbol{-1.73}[-2.53,-1.21]$ & $0.42[-0.39,1.18]$ & $0.08[-0.47,0.59]$ & $0.62[0.54,0.70]$ \\
    \bottomrule
  \end{tabular}
\end{table}
 
\subsection{Minesweeper}\label{supp:feature-results-minesweeper}
\label{app:minesweeper-features}

Let $V_t$ be the hidden cells and $C_t$ the displayed numbered clues.
Associate the binary variable $X_v$ with each $v\in V_t$, where one denotes a
mine.
Displayed flags are fixed known mines, not variables.
For clue $a\in C_t$, let $d_t(a)$ be its displayed value and $N_t(a)$
its neighboring cells. Define the residual mine count
\begin{equation}\label{eq:ms-residual-mines}
  r_t(a)
  =d_t(a)-\sum_{f\in N_t(a)}
    \mathbf 1\{f\text{ is a displayed flag}\}.
\end{equation}
The corresponding clue equation is
\begin{equation}
  \sum_{v\in N_t(a)\cap V_t}X_v=r_t(a).
  \label{eq:minesweeper-clue-equation}
\end{equation}
The task imposes no additional constraint on the board's total mine count.

For $Q\subseteq C_t$, let
\begin{equation}
  \mathcal A_t(Q)
  =\left\{\boldsymbol{z}\in\{0,1\}^{V_t}:
    \sum_{v\in N_t(a)\cap V_t}z_v=r_t(a)
    \text{ for every }a\in Q\right\}.
  \label{eq:minesweeper-subset-assignments}
\end{equation}
Candidate $j$ is the cell $v_{tj}$, and $b_t\in\{0,1\}$ is the requested
status, safe or mine.
A clue subset $Q$ certifies $j$ when
$z_{v_{tj}}=b_t$ for every $\boldsymbol{z}\in\mathcal A_t(Q)$.
Every such subset is satisfiable because the displayed puzzle has a globally
consistent assignment.

\paragraph{Minimum certificate size.}
Define
$S_{tj}=\min\{|Q|:Q\subseteq C_t,\ Q\text{ certifies }j\}$.
This is the fewest displayed clue equations sufficient to force the requested
status of the candidate cell.

\paragraph{Minimum certificate scope.}
For a clue subset $Q$, define
\begin{equation}
\begin{aligned}
  \operatorname{vars}_t(Q)
    &=\bigcup_{a\in Q}\bigl(N_t(a)\cap V_t\bigr),\\
  \mathcal Q^\star_{tj}
    &=\{Q\subseteq C_t:Q\text{ certifies }j,\ |Q|=S_{tj}\},\\
  B_{tj}
    &=\min_{Q\in\mathcal Q^\star_{tj}}|\operatorname{vars}_t(Q)|.
\end{aligned}
\label{eq:minesweeper-certificate-scope}
\end{equation}
The lexicographic construction first minimizes the number of clues and then
the number of hidden-cell states touched by a shortest certificate.
The features are computed exactly by enumerating hidden-cell assignments and
testing every displayed-clue subset.
Raw certificate size ranges from one to four; raw scope ranges from one to
nine.
\begin{table}[H]
  \caption{Selected-solution Minesweeper regressions.}
  \label{tab:minesweeper-choice-results}
  \centering
  \regressiontableformat
  \begin{tabular}{@{}lrrr@{}}
    \toprule
    Source & $\widehat{\theta}_{\mathrm{size}}$ & $\widehat{\theta}_{\mathrm{scope}}$ & Acc. \\
    \midrule
    Human & $\boldsymbol{-0.48}[-0.87,-0.09]$ & $-0.17[-0.58,0.18]$ & $0.29[0.21,0.34]$ \\
    GPT & $0.34[-1.08,2.19]$ & $-0.06[-1.98,1.36]$ & $0.21[0.04,0.40]$ \\
    Claude & $-0.66[-2.32,1.09]$ & $-0.59[-2.15,0.45]$ & $0.32[0.17,0.40]$ \\
    Gemini & $-1.22[-3.53,0.95]$ & $-0.46[-2.40,0.66]$ & $0.34[0.12,0.47]$ \\
    \bottomrule
  \end{tabular}
\end{table}

\begin{table}[H]
  \caption{Source classification for Minesweeper.}
  \label{tab:minesweeper-source-binary-results}
  \centering
  \regressiontableformat
  \begin{tabular}{@{}lrrr@{}}
    \toprule
    Model/Human & $\widehat{\beta}_{\mathrm{size}}$ & $\widehat{\beta}_{\mathrm{scope}}$ & BA \\
    \midrule
    GPT & $0.74[-0.22,2.25]$ & $0.12[-1.35,1.01]$ & $0.56[0.38,0.64]$ \\
    Claude & $-0.17[-1.35,0.98]$ & $-0.36[-1.38,0.35]$ & $0.53[0.43,0.57]$ \\
    Gemini & $-0.66[-2.10,0.85]$ & $-0.25[-1.68,0.42]$ & $0.52[0.44,0.59]$ \\
    \bottomrule
  \end{tabular}
\end{table}
 
\subsection{Sudoku}\label{supp:feature-results-sudoku}
\label{app:sudoku-features}

Let $n_t\in\{4,6\}$ be the grid side length, let $d_t\in[n_t]$ be the named
target digit, and let $v_{tj}$ be the empty cell represented by candidate
$j\in\mathcal J_t$.
Write
$\mathcal U_t(v)=\{u_t^{\mathsf{row}}(v),
u_t^{\mathsf{col}}(v),u_t^{\mathsf{box}}(v)\}$
for the three units containing cell $v$.

\paragraph{Cell-domain size.}
For an empty cell $v$, define
\begin{equation}
  \mathcal D_t(v)
  =\{d\in[n_t]:d\text{ is absent from every }
    u\in\mathcal U_t(v)\},
  \qquad
  D_{tj}=|\mathcal D_t(v_{tj})|.
  \label{eq:sudoku-domain-size}
\end{equation}
A smaller value means less other-digit ambiguity at the candidate cell;
$D_{tj}=1$ is the cell-based, naked-single-like extreme.

\paragraph{Target-location count.}
For a unit $u\in\mathcal U_t(v_{tj})$, define
\begin{equation}
  \lambda_{tju}
  =\sum_{\ell\in\mathcal J_t}
    \mathbf 1\{v_{t\ell}\in u\},
  \qquad
  T_{tj}
  =\min_{u\in\mathcal U_t(v_{tj})}\lambda_{tju}.
  \label{eq:sudoku-target-location-count}
\end{equation}
This is the smallest number of currently legal locations for the target digit
among the candidate's row, column, and box.
A value of one is the digit-based, hidden-single-like extreme.

\FloatBarrier
\begin{table}[H]
  \caption{Selected-solution Sudoku regressions.}
  \label{tab:sudoku-choice-results}
  \centering
  \regressiontableformat
  \begin{tabular}{@{}lrrr@{}}
    \toprule
    Source & $\widehat{\theta}_{\mathrm{domain}}$ & $\widehat{\theta}_{\mathrm{location}}$ & Acc. \\
    \midrule
    Human & $-0.20[-0.53,0.08]$ & $\boldsymbol{-0.89}[-1.12,-0.68]$ & $0.28[0.21,0.33]$ \\
    GPT & $-0.40[-1.86,0.97]$ & $\boldsymbol{-1.06}[-4.98,-0.27]$ & $0.28[0.03,0.45]$ \\
    Claude & $0.11[-0.97,1.26]$ & $\boldsymbol{-0.81}[-1.83,-0.16]$ & $0.10[0.06,0.41]$ \\
    Gemini & $-0.42[-1.32,0.37]$ & $\boldsymbol{-1.21}[-3.20,-0.61]$ & $0.28[0.08,0.39]$ \\
    \bottomrule
  \end{tabular}
\end{table}

\begin{table}[H]
  \caption{Source classification for Sudoku.}
  \label{tab:sudoku-source-binary-results}
  \centering
  \regressiontableformat
  \begin{tabular}{@{}lrrr@{}}
    \toprule
    Model/Human & $\widehat{\beta}_{\mathrm{domain}}$ & $\widehat{\beta}_{\mathrm{location}}$ & BA \\
    \midrule
    GPT & $-0.16[-1.39,0.84]$ & $-0.11[-1.55,0.43]$ & $0.49[0.33,0.61]$ \\
    Claude & $0.25[-0.55,1.07]$ & $0.06[-0.64,0.49]$ & $0.51[0.37,0.58]$ \\
    Gemini & $-0.19[-0.76,0.30]$ & $-0.22[-1.43,0.15]$ & $0.53[0.42,0.58]$ \\
    \bottomrule
  \end{tabular}
\end{table}
 
\FloatBarrier
\Needspace{46\baselineskip}
\section{Perturbation study}\label{supp:perturbation-study}
\label{supp:target-comparisons}
\label{supp:perturbation-bootstrap}\label{sec:distribution-tests}

This section gives family-level results for Figure~\ref{fig:presentation}.
Perturbations are described in Section~\ref{supp:puzzle-perturbations}
and measures in Methods, Section~\ref{sec:target-methods}.
\paragraph{Confidence intervals.}
We use 5,000 whole-puzzle bootstrap resamples within family, keeping each
puzzle's comparisons together and empirical response distributions 
for each  task fixed
(Section~\ref{sec:methods-statistics}). Normalized-change intervals use
eligible puzzles; related-minus-unrelated TV and MI
differences between contexts or sources use matched puzzle differences

\paragraph{Permutation tests.}
For each test, we compare the observed statistic with values calculated after repeatedly shuffling the data as described below. To test equality of answer distributions across variants, variant labels are permuted among the combined answers from the two variants while preserving sample sizes. To test independence of paired answers, context-puzzle answers are held fixed and target-puzzle answers are permuted across pairs, preserving both marginal distributions. Permutations are performed separately within each comparison. For statistics whose larger values indicate greater departure from the null, such as total variation distance or mutual information, we calculate $p=(1+N_{\ge})/(B+1)$, where $B$ is the number of permutations and $N_{\ge}$ counts permuted statistics at least as large as observed.

TV tests use 10,000 random permutations, with Holm adjustment across all 320 comparisons shown in Table~\ref{tab:perturbation_tests}.

The individual-puzzle MI tests reported here use 2,000 permutations. Their 80 unadjusted $p$-values, covering ten puzzles, two contexts and four sources, are summarized in Table~\ref{tab:context-mi-mean-p}; individual results are provided in the source data.

\begin{table}[H]
\centering
\caption{\textbf{TV permutation tests.} Cells report mean unadjusted $p$, followed by the number of comparisons with Holm-adjusted $p\le0.05$ out of the total.}
\label{tab:perturbation_tests}
\small
\setlength{\tabcolsep}{6pt}
\begin{tabular}{@{}lcccc@{}}
\toprule
Module & Human & GPT & Claude & Gemini \\
\midrule
Highlighting & 0.0917 (10/20) & 0.3034 (10/20) & 0.0146 (13/20) & 0.0935 (13/20) \\
Related context & 0.0010 (9/10) & 0.8383 (1/10) & 0.3363 (3/10) & 0.1878 (2/10) \\
Strategy primer & 0.2549 (0/10) & 0.6074 (2/10) & 0.3044 (3/10) & 0.2272 (3/10) \\
Spatial reflection & 0.1103 (6/30) & 0.0337 (28/30) & 0.0133 (28/30) & 0.0255 (28/30) \\
Number ordering & 0.2332 (1/10) & 0.1525 (5/10) & 0.0424 (4/10) & 0.1556 (2/10) \\
\bottomrule
\end{tabular}
\end{table}
 \FloatBarrier
\Needspace{32\baselineskip}
\subsection{Highlighting}\label{supp:perturbation-results-highlighting}
Much of the observed TV reflects increased probability of highlighted
solutions. TV tests detect changes from the unhighlighted reference for
humans and each model, though not in every comparison.

\begin{table}[H]
  \caption{Highlighting. TV is $d_{\mathrm{TV}}$ from the unhighlighted reference; probability changes concern the highlighted solution set (Section~\ref{app:stimulus-measures}). Entries average arms within puzzles. Mean weights families equally; brackets show 95\% puzzle-bootstrap CIs.}
  \label{tab:cue-detail}
  \centering
  \setlength{\tabcolsep}{2pt}
  \renewcommand{\arraystretch}{1.2}
  \begin{adjustbox}{max width=\linewidth}
  \begin{tabular}{@{}lcccc@{}}
    \toprule
    Quantity & Human & GPT & Claude & Gemini \\
    \midrule
    \multicolumn{5}{l}{\textit{Minesweeper}} \\
    Prob. change & $0.19[0.16,0.21]$ & $0.50[0.50,0.50]$ & $0.50[0.50,0.50]$ & $0.49[0.48,0.50]$ \\
    Normalized prob. change & $0.33[0.27,0.39]$ & $1.00[0.99,1.00]$ & $0.88[0.68,1.00]$ & $0.97[0.93,1.00]$ \\
    TV from ref. & $0.20[0.18,0.23]$ & $0.53[0.51,0.55]$ & $0.52[0.50,0.54]$ & $0.52[0.49,0.55]$ \\
    \midrule
    \multicolumn{5}{l}{\textit{Sudoku}} \\
    Prob. change & $0.27[0.25,0.30]$ & $0.50[0.49,0.50]$ & $0.50[0.50,0.50]$ & $0.50[0.49,0.50]$ \\
    Normalized prob. change & $0.55[0.49,0.61]$ & $0.99[0.98,1.00]$ & $1.00[1.00,1.00]$ & $0.99[0.98,1.00]$ \\
    TV from ref. & $0.29[0.27,0.32]$ & $0.50[0.49,0.50]$ & $0.55[0.53,0.57]$ & $0.53[0.51,0.55]$ \\
    \midrule
    \multicolumn{5}{l}{\textit{Mean}} \\
    Prob. change & $0.23[0.21,0.25]$ & $0.50[0.49,0.50]$ & $0.50[0.50,0.50]$ & $0.49[0.48,0.50]$ \\
    Normalized prob. change & $0.44[0.40,0.48]$ & $1.00[0.99,1.00]$ & $0.94[0.84,1.00]$ & $0.98[0.96,1.00]$ \\
    TV from ref. & $0.25[0.23,0.27]$ & $0.51[0.50,0.53]$ & $0.54[0.52,0.55]$ & $0.52[0.50,0.54]$ \\
    \bottomrule
  \end{tabular}
  \end{adjustbox}
\end{table}
 \FloatBarrier

\Needspace{34\baselineskip}
\subsection{Spatial reflection}\label{supp:perturbation-results-reflection}
Model responses are particularly sensitive to top--bottom reflection.

\begin{table}[H]
  \caption{Spatial reflection. TV from ref. is $d_{\mathrm{TV}}$ from the reference presentation, after mapping answers back to their original cells. Mean rows average the three reflections; the Mean group weights families equally. Brackets show 95\% puzzle-bootstrap CIs.}
  \label{tab:spatial-detail}
  \centering
  \setlength{\tabcolsep}{2pt}
  \renewcommand{\arraystretch}{1.2}
  \begin{adjustbox}{max width=\linewidth}
  \begin{tabular}{@{}lcccc@{}}
    \toprule
    Quantity & Human & GPT & Claude & Gemini \\
    \midrule
    \multicolumn{5}{l}{\textit{Minesweeper}} \\
    Left--right & $0.14[0.12,0.16]$ & $0.52[0.19,0.86]$ & $0.45[0.22,0.67]$ & $0.25[0.06,0.48]$ \\
    Top--bottom & $0.20[0.15,0.26]$ & $1.00[0.99,1.00]$ & $0.80[0.75,0.83]$ & $0.66[0.44,0.89]$ \\
    Both & $0.20[0.11,0.32]$ & $0.99[0.99,1.00]$ & $0.79[0.67,0.90]$ & $0.67[0.40,0.90]$ \\
    Mean & $0.18[0.14,0.23]$ & $0.84[0.73,0.95]$ & $0.68[0.58,0.77]$ & $0.53[0.34,0.72]$ \\
    \midrule
    \multicolumn{5}{l}{\textit{Sudoku}} \\
    Left--right & $0.22[0.17,0.30]$ & $0.99[0.98,1.00]$ & $0.93[0.90,0.96]$ & $0.87[0.81,0.92]$ \\
    Top--bottom & $0.23[0.19,0.28]$ & $1.00[1.00,1.00]$ & $0.91[0.88,0.94]$ & $0.85[0.67,0.95]$ \\
    Both & $0.26[0.23,0.29]$ & $1.00[1.00,1.00]$ & $0.94[0.92,0.97]$ & $0.80[0.65,0.92]$ \\
    Mean & $0.24[0.21,0.26]$ & $1.00[0.99,1.00]$ & $0.93[0.92,0.94]$ & $0.84[0.71,0.92]$ \\
    \midrule
    \multicolumn{5}{l}{\textit{Mean}} \\
    Left--right & $0.18[0.15,0.22]$ & $0.76[0.59,0.93]$ & $0.69[0.57,0.80]$ & $0.56[0.46,0.68]$ \\
    Top--bottom & $0.21[0.18,0.25]$ & $1.00[1.00,1.00]$ & $0.86[0.83,0.88]$ & $0.75[0.61,0.88]$ \\
    Both & $0.23[0.18,0.29]$ & $1.00[0.99,1.00]$ & $0.87[0.81,0.92]$ & $0.73[0.58,0.87]$ \\
    Mean & $0.21[0.18,0.24]$ & $0.92[0.86,0.97]$ & $0.80[0.75,0.85]$ & $0.68[0.57,0.79]$ \\
    \bottomrule
  \end{tabular}
  \end{adjustbox}
\end{table}
 
\FloatBarrier
\Needspace{46\baselineskip}
\subsection{Preceding-puzzle context}\label{supp:perturbation-results-context}
The mean related-minus-unrelated MI difference is largest for humans,
driven by Minesweeper; the human Sudoku interval includes zero.
Holm-adjusted TV tests reject independence in nine of ten related human
comparisons, versus at most three for AI model.

\FloatBarrier

\begin{table}[H]
  \caption{Preceding-puzzle context by family. Measures are defined in Methods, Section~\ref{sec:target-methods}; $\Delta$ denotes related minus unrelated context. Estimates use pairs with two valid answers from five puzzles per family. Mean weights families equally. Brackets show 95\% whole-puzzle bootstrap intervals.}
  \label{tab:pair-movement}
  \centering
  \setlength{\tabcolsep}{2pt}
  \renewcommand{\arraystretch}{1.2}
  \begin{adjustbox}{max width=\linewidth}
  \begin{tabular}{@{}lcccc@{}}
    \toprule
    Quantity & Human & GPT & Claude & Gemini \\
    \midrule
    \multicolumn{5}{l}{\textit{Minesweeper}} \\
    Related TV from ref. & $0.36[0.32,0.42]$ & $0.00[0.00,0.01]$ & $0.13[0.04,0.23]$ & $0.12[0.03,0.22]$ \\
    Unrelated TV from ref. & $0.25[0.23,0.28]$ & $0.06[0.00,0.14]$ & $0.03[0.01,0.05]$ & $0.08[0.01,0.17]$ \\
    $\Delta\mathrm{TV}$ & $0.11[0.05,0.17]$ & $-0.05[-0.12,0.00]$ & $0.10[0.01,0.21]$ & $0.05[0.00,0.12]$ \\
    \shortstack[l]{Normalized prob. change\\(related)} & $0.30[0.19,0.41]$ & $0.00[0.00,0.00]$ & $0.10[0.01,0.20]$ & $0.08[0.00,0.18]$ \\
    Related MI & $0.61[0.52,0.77]$ & $0.01[0.00,0.02]$ & $0.16[0.04,0.30]$ & $0.17[0.04,0.31]$ \\
    Unrelated MI & $0.35[0.30,0.40]$ & $0.06[0.00,0.14]$ & $0.03[0.01,0.06]$ & $0.09[0.01,0.18]$ \\
    $\Delta I$ & $0.26[0.14,0.43]$ & $-0.06[-0.12,0.00]$ & $0.13[0.01,0.28]$ & $0.08[0.00,0.17]$ \\
    \midrule
    \multicolumn{5}{l}{\textit{Sudoku}} \\
    Related TV from ref. & $0.28[0.25,0.31]$ & $0.07[0.00,0.22]$ & $0.08[0.01,0.16]$ & $0.11[0.07,0.16]$ \\
    Unrelated TV from ref. & $0.28[0.25,0.32]$ & $0.00[0.00,0.00]$ & $0.10[0.03,0.18]$ & $0.09[0.04,0.17]$ \\
    $\Delta\mathrm{TV}$ & $0.01[-0.03,0.06]$ & $0.07[0.00,0.22]$ & $-0.03[-0.09,0.02]$ & $0.02[-0.06,0.11]$ \\
    \shortstack[l]{Normalized prob. change\\(related)} & $0.20[0.14,0.26]$ & $0.05[0.00,0.16]$ & $0.05[-0.01,0.15]$ & $0.05[0.02,0.09]$ \\
    Related MI & $0.43[0.31,0.56]$ & $0.10[0.00,0.29]$ & $0.11[0.02,0.21]$ & $0.12[0.05,0.20]$ \\
    Unrelated MI & $0.45[0.34,0.57]$ & $0.00[0.00,0.00]$ & $0.13[0.04,0.21]$ & $0.09[0.04,0.17]$ \\
    $\Delta I$ & $-0.02[-0.13,0.12]$ & $0.10[0.00,0.29]$ & $-0.02[-0.12,0.09]$ & $0.03[-0.04,0.13]$ \\
    \midrule
    \multicolumn{5}{l}{\textit{Mean}} \\
    Related TV from ref. & $0.32[0.30,0.35]$ & $0.04[0.00,0.11]$ & $0.10[0.04,0.17]$ & $0.12[0.07,0.17]$ \\
    Unrelated TV from ref. & $0.26[0.24,0.29]$ & $0.03[0.00,0.07]$ & $0.07[0.03,0.11]$ & $0.09[0.04,0.14]$ \\
    $\Delta\mathrm{TV}$ & $0.06[0.02,0.10]$ & $0.01[-0.05,0.09]$ & $0.04[-0.02,0.10]$ & $0.03[-0.02,0.09]$ \\
    \shortstack[l]{Normalized prob. change\\(related)} & $0.25[0.19,0.32]$ & $0.03[0.00,0.08]$ & $0.08[0.01,0.14]$ & $0.07[0.02,0.12]$ \\
    Related MI & $0.52[0.43,0.62]$ & $0.05[0.00,0.15]$ & $0.13[0.06,0.22]$ & $0.15[0.07,0.23]$ \\
    Unrelated MI & $0.40[0.34,0.47]$ & $0.03[0.00,0.07]$ & $0.08[0.04,0.12]$ & $0.09[0.04,0.15]$ \\
    $\Delta I$ & $0.12[0.03,0.23]$ & $0.02[-0.05,0.13]$ & $0.05[-0.03,0.14]$ & $0.06[0.00,0.12]$ \\
    \bottomrule
  \end{tabular}
  \end{adjustbox}
\end{table}
 \FloatBarrier

\Needspace{40\baselineskip}
\begin{figure}[H]
\centering
\includegraphics[width=\linewidth]{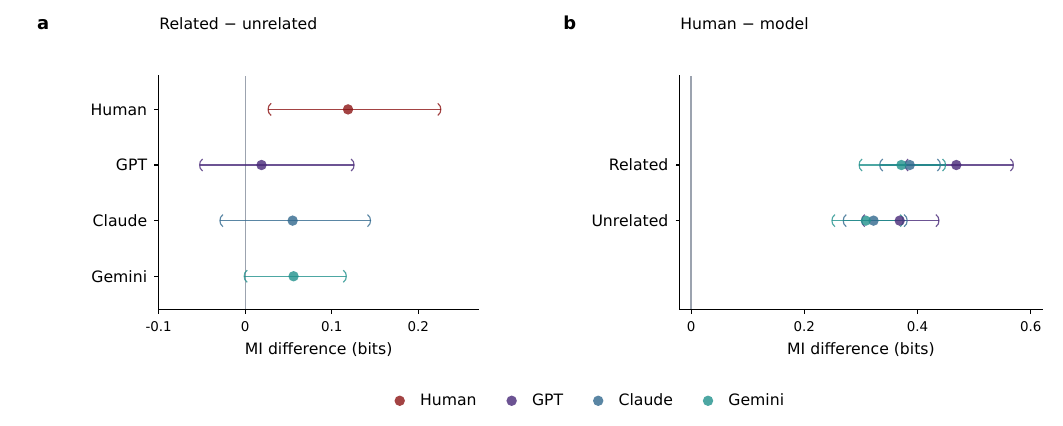}
\caption{\textbf{Differences in mutual information across contexts and sources.}
\textbf{a}, Related minus unrelated MI for each source.
\textbf{b}, Human minus model MI in each context. Points are equal-family
mean differences over five Minesweeper and five Sudoku puzzles; bars are
pointwise 95\% bootstrap intervals (Section~\ref{supp:perturbation-bootstrap}).
Estimates use pairs with two valid answers; models use low effort and plain prompts.}
\label{fig:context-mi-contrasts}
\end{figure}

The related-minus-unrelated MI difference is positive for humans, with a 95\% bootstrap interval excluding zero; the corresponding intervals include zero for all three AI models (Figure~\ref{fig:context-mi-contrasts}a). Human-minus-model MI differences are positive in both contexts, with all six 95\% bootstrap intervals excluding zero (Figure~\ref{fig:context-mi-contrasts}b).

\begin{table}[H]
\centering
\caption{\textbf{Individual-puzzle MI permutation tests.} Mean $p$ averages ten unadjusted $p$-values per source and context (five puzzles per family); counts give the number below 0.05. Mean $p$ is descriptive, not a combined $p$-value or a test of mean MI. Models use low effort and plain prompts; tests are described in Section~\ref{supp:perturbation-bootstrap}.}
\label{tab:context-mi-mean-p}
\setlength{\tabcolsep}{10pt}
\begin{tabular}{@{}lcccc@{}}
\toprule
& \multicolumn{2}{c}{Related} & \multicolumn{2}{c}{Unrelated} \\
\cmidrule(lr){2-3}\cmidrule(l){4-5}
Source & Mean $p$ & $p<0.05$ & Mean $p$ & $p<0.05$ \\
\midrule
Human & 0.0086 & 9/10 & 0.0083 & 10/10 \\
GPT & 0.8381 & 1/10 & 0.8002 & 2/10 \\
Claude & 0.3297 & 6/10 & 0.5391 & 2/10 \\
Gemini & 0.2254 & 6/10 & 0.3830 & 5/10 \\
\bottomrule
\end{tabular}
\end{table}
 \FloatBarrier

\Needspace{32\baselineskip}
\subsection{Strategy primer}\label{supp:perturbation-results-primer}
For every source, the 95\% interval for the equal-family mean normalized
probability change includes zero (Figure~\ref{fig:presentation}b).
TV tests detect changes from the no-primer reference in some model
comparisons and none of the human comparisons. These changes need not
favor the primer-matching answer.

\begin{table}[H]
  \caption{Strategy-primer distribution and probability changes. TV is $d_{\mathrm{TV}}$ from the no-primer reference. Probability changes concern the designated solution (Section~\ref{app:stimulus-measures}). Mean weights families equally; brackets show 95\% puzzle-bootstrap CIs.}
  \label{tab:transfer-attraction}
  \label{tab:transfer-movement}
  \centering
  \setlength{\tabcolsep}{2pt}
  \renewcommand{\arraystretch}{1.2}
  \begin{adjustbox}{max width=\linewidth}
  \begin{tabular}{@{}lcccc@{}}
    \toprule
    Quantity & Human & GPT & Claude & Gemini \\
    \midrule
    \multicolumn{5}{l}{\textit{Minesweeper}} \\
    TV from ref. & $0.09[0.06,0.11]$ & $0.16[0.01,0.39]$ & $0.14[0.03,0.25]$ & $0.24[0.07,0.46]$ \\
    Prob. change & $0.02[-0.04,0.07]$ & $0.02[-0.03,0.10]$ & $-0.05[-0.17,0.04]$ & $0.19[0.03,0.41]$ \\
    Normalized prob. change & $0.03[-0.06,0.11]$ & $0.02[-0.03,0.10]$ & $-0.10[-0.32,0.04]$ & $0.18[0.01,0.41]$ \\
    \midrule
    \multicolumn{5}{l}{\textit{Sudoku}} \\
    TV from ref. & $0.08[0.05,0.11]$ & $0.01[0.00,0.04]$ & $0.12[0.07,0.20]$ & $0.11[0.04,0.20]$ \\
    Prob. change & $-0.01[-0.05,0.04]$ & $0.00[0.00,0.00]$ & $0.03[-0.09,0.16]$ & $-0.05[-0.17,0.05]$ \\
    Normalized prob. change & $-0.07[-0.17,0.03]$ & $0.00[0.00,0.00]$ & $-0.24[-1.38,0.50]$ & $-0.29[-1.19,0.35]$ \\
    \midrule
    \multicolumn{5}{l}{\textit{Mean}} \\
    TV from ref. & $0.08[0.06,0.10]$ & $0.09[0.01,0.20]$ & $0.13[0.07,0.20]$ & $0.18[0.08,0.29]$ \\
    Prob. change & $0.00[-0.04,0.04]$ & $0.01[-0.01,0.05]$ & $-0.01[-0.09,0.07]$ & $0.07[-0.03,0.19]$ \\
    Normalized prob. change & $-0.02[-0.09,0.04]$ & $0.01[-0.01,0.05]$ & $-0.17[-0.74,0.23]$ & $-0.05[-0.51,0.28]$ \\
    \bottomrule
  \end{tabular}
  \end{adjustbox}
\end{table}
 \FloatBarrier
 
\begingroup\small\raggedright
+\newcommand{\etalchar}[1]{$^{#1}$}
\providecommand{\bysame}{\leavevmode\hbox to3em{\hrulefill}\thinspace}
\providecommand{\MR}{\relax\ifhmode\unskip\space\fi MR }
\providecommand{\MRhref}[2]{%
  \href{http://www.ams.org/mathscinet-getitem?mr=#1}{#2}
}
\providecommand{\href}[2]{#2}

\endgroup
\end{cbunit}
\clearpage{}
\end{document}